\documentclass[11pt]{article}
\usepackage{epsfig,graphics,bm}

\renewcommand{\baselinestretch}{1.0}

\newcommand{\mar}[1]{\marginpar{\vspace{15pt}\textsf{\footnotesize #1}}}

\def\spose#1{\hbox to 0pt{#1\hss}}
\def\gta{\mathrel{\spose{\lower 3pt\hbox{$\mathchar"218$}}
     \raise 2.0pt\hbox{$\mathchar"13E$}}}

\begin{document}
\pagestyle{plain}

\section*{\centering \sffamily
Modeling the 3--D Secular Planetary Three-Body Problem. Discussion on the outer  ${\upsilon}$ Andromedae Planetary System} \vspace{1cm}

\begin{center}
T.A. Michtchenko, S. Ferraz-Mello,

Instituto de Astronomia, Geof\'{\i}sica e Ci\^encias Atmosf\'ericas, USP\\
Rua do Mat\~ao 1226, 05508-900 S\~ao Paulo, Brazil

E-mail: tatiana@astro.iag.usp.br

\vspace{0.4cm} {and C. Beaug\'{e}}

Observatorio Astron\'omico, Universidad Nacional de C\'ordoba,
Laprida 854, (X5000BGR) C\'ordoba, Argentina

\end{center}

\vfill
\begin{itemize}
\item Total number of pages: 37
\item Number of manuscript pages: 24
\item Number of figures: 10
\item Number of tables: 1
\end{itemize}

\newpage
\noindent {\bf Proposed running head:} Modeling the 3--D Secular Planetary Three-Body Problem.

\vspace{2cm}

{\bf Editorial correspondence to:}\\
 Tatiana A. Michtchenko\\
 Instituto de Astronomia, Geof\a'{\i}sica e Ci\^encias Atmosf\a'ericas\\
 Universidade de S\~ao Paulo\\
 Rua do Mat\~ao, 1226\\
 S\~ao Paulo, Brasil 05508-900\\
 Email: tatiana@astro.iag.usp.br

\newpage

\renewcommand{\baselinestretch}{1.0}
\normalsize

\begin{abstract}
The three-dimensional secular behavior of a system composed of a central star and two massive planets is modeled semi-analytically in the frame of the general three-body problem. The main dynamical features of the system are presented in geometrical pictures allowing us to investigate a large domain of the phase space of this problem without time-expensive numerical integrations of the equations of motion and without any restriction on the magnitude of the planetary eccentricities, inclinations and mutual distance. Several regimes of motion of the system are observed. With respect to the secular angle $\Delta\varpi$, possible motions are circulations, oscillations (around $0$ and $180^\circ$), and high eccentricity/inclination librations in secular resonances. With respect to the arguments of pericenter, $\omega_1$ and $\omega_2$, possible motions are direct circulation and high-inclination libration around $\pm 90^\circ$ in the Lidov-Kozai resonance. The regions of transition between domains of different regimes of motion are characterized by chaotic behavior.

We apply the analysis to the case of the two outer planets of the $\upsilon$ Andromedae system, observed edge-on. The topology of the 3--D phase space of this system is investigated in detail by means of surfaces of section, periodic orbits and dynamical spectra, mapping techniques and numerical simulations. We obtain the general structure of the phase space, and the boundaries of the spatial secular stability. We find that this system is secularly stable in a large domain of eccentricities and inclinations.

\end{abstract}

\vspace{1cm}

\noindent
Key words: Celestial mechanics; Extrasolar planets; Planetary dynamics;
Resonances

\newpage

\renewcommand{\baselinestretch}{1.0}  \normalsize

\section{Introduction}

The discovery of extra-solar planets has brought into focus a long standing, but yet unsolved problem in Celestial Mechanics, namely the long-term stability of planetary systems. In fact, the dynamical evolution of the Solar System has been the object of exhaustive studies for at least two centuries, but its stability is not yet completely understood. The extra-solar systems provoke our imagination by their unusual configurations, which include Jupiter-mass planets at very small semi-major axes and unexpectedly large eccentricities, and raise new questions in our understanding of their dynamical stability.

The first studies on the long-term stability of extra-solar planetary systems have been done by means of direct numerical integrations of the full equations of motion. In the last years, a large number of papers were published applying such approach to several planetary systems, in order to investigate their dynamical aspects. The numerical integration over very-long time intervals is a powerful approach for exploring the dynamical features of the systems with two and more planets. However, the correct interpretation of the results of numerical investigations is not simple and should be always founded on the basic concepts provided by analytical or semi-analytical investigations. Only these approaches are able to explore the whole parameter space of the problem and provide the decisive answer on the question of the global stability of the system under study.

Until recently, the analytical models founded on the Lagrange-Laplace secular perturbation theory were commonly used in the study of the secular three-dimensional dynamics of the general three-body problem. These models are based on the classical expansion of the disturbing function in powers of eccentricities and inclinations, therefore their application is unsurprisingly restricted to the planetary systems with small eccentricities and inclinations (Murray and Dermott 1999). Even when a large number of the terms in the disturbing function is retained (Laskar 1987, Laskar and Robutel 1995), the convergence of the classical expansion remains limited to small values of the eccentricities, in accordance with the Sundman criterion (see Ferraz-Mello 1994), which establishes the convergence domains in the space of eccentricities for a given ratio of the semi-major axes.

Analytical approaches based on the expansion of the disturbing function in a power series of the ratio of the planetary semi-major axes, $\alpha=a_1/a_2$, have been used for triple star systems (Harrington 1969, Ford {\it et al}. 2000). The expansion to order $\alpha^3$, called octopole approximation, has been recently applied to study the secular dynamics of systems with two planets (Lee and Peale 2003). However, since the convergence of the series is generally very slow, the applicability of this approach is limited to satellite studies or, in planetary dynamics, to the domain of the so-called "hierarchical systems", which are characterized by suitably small values of $\alpha$.

In a preceding paper (Michtchenko and Malhotra 2004, hereafter referred to as MM2004), we introduced a new semi-analytical approach, consisting of a numerical averaging over the short-period perturbations in the mutual interaction of the two planets on coplanar orbits. We obtained the planar secular Hamiltonian that allowed us to overcome the limitations of the classical theories. Our approach revealed to be suitable for the study of the numerous extra-solar systems whose orbital characteristics (e.g. eccentricities and mutual distances) are beyond the range of the classical theories. We have investigated the topology of the phase space of the two (coplanar) outer planets, {\bf c} and {\bf d}, of $\upsilon$ Andromedae system  (Butler {\it et al.} 1997, Butler {\it et al.} 1999). Our study permitted to explain such common features of planetary systems as the coupled apsidal line motion with aligned and anti-aligned pericenters. In addition, the analysis allowed us to discover new dynamical features in planetary systems, namely the existence of a nonlinear secular resonance in the very-high-eccentricity region.

The goal of the present paper is to extend that study to the non-coplanar secular motion of the planets. It is known that, at present, the majority of the discovered extra-solar systems are spatially unresolved. In fact, the spectroscopic radial velocity technique measures only the line-of-sight component of the velocity of the star and the fitting of Keplerian ellipses is unable to detect the inclinations and nodes of the planetary orbits. A system must be observed over a large number of orbital periods, in order to allow a full N-body fit able to determine all the orbital parameters of the system (Ferraz-Mello {\it et al.} 2005). The exceptions are the multi-planet systems of strongly interacting companions involved in mean-motion resonances. Presently, only the two planets, {\bf b} and {\bf c}, around PSR B1257+12, which are near the 3:2 mean-motion resonance, have the individual orbital inclinations dynamically determined (Konacki and Wolszczan 2003).

Although the spectroscopic observational data do not constrain the orbital inclinations, dynamical stability considerations can still provide constraints on the individual planetary masses and inclinations. For instance, for the $\upsilon$ Andromedae system, dynamical arguments suggest that the mutual inclination between planetary orbits is less than $20^\circ$ and that the mean inclination to the plane of the sky is greater than $13^\circ$  (Stepinski {\it et al.} 2000, Chiang {\it et al.} 2001, Lissauer and Rivera 2001). These limits are consistent with the results of the analysis of {\it Hipparcos} astrometric measurements, yielding an estimate of the true mass of the outermost companion and constraining its orbital inclination to the line of sight to be greater than $11^\circ$ (Mazeh {\it et al.} 1999). Also, no transits of the innermost planet were found in the photometry observations, limiting its orbital inclination to less than $83^\circ$ (Lissauer 1999). 

It is expected that astrometric and photometric techniques contribute to the task of complete characterization of known systems (e.g. the masses and the spatially resolved orbital elements). In analogy with the diversity observed in the planar orbital parameters of the known extra-solar systems, the elements to be determined may show spatial characteristics that are hard to explain within the context of current theoretical models based on the Laplace-Lagrange secular theory (Veras and Armitage 2004). We hope that this paper may help to face future challenges.  

The non-linear model developed in this work describes the three-dimensional motion of the two planets revolving around the same star. It is based  on the same semi-analytical approach used in MM2004, consequently, it preserves all advantages of the 2--D model. The model is applied to the study of the topology of the 3--D phase space of one planetary system with the dynamical parameters of the $\upsilon$ Andromedae (planets {\bf c} and {\bf d}). The results, presented and discussed in detail in Sections 4 and 5, are in good accordance with those obtained through direct numerical integration of equations of motion. This is mainly due to the fact that the model presented here has no restrictions on the magnitude of the planet eccentricities, inclinations and on the planetary spacing. Moreover, the present model allows us to characterize the phase space of a given system through the calculation of energy levels and families of periodic orbits, independently of high-cost direct integration of the exact equations of the motion. The results are completed with direct integrations; it is worth emphasizing that the combined use of the two distinct approaches shows very good agreement.

The paper is organized as follows. In section 2, we introduce the 3--D model and describe the basic concepts of the 3--D general three-body problem. They allow us to reduce a number of degrees of freedom of the Hamiltonian system to 2. In section 3, we show that the secular behavior of the system can be characterized through the geometrical pictures of its phase space. The important point is that their construction does not involve the time-consuming numerical integrations. In sections 4 and 5, we apply our theory to the outer $\upsilon$ Andromedae system,  assuming the edge-on observed planets {\bf c} and {\bf d}. The topology and dynamical maps of the totally reduced (to 2 degrees of freedom) system on the representative subspaces of initial conditions are presented in section 4, together with the study of the dependence of the secular dynamics on the magnitude of the Angular Momentum Deficit. In section 5, we study the partially reduced (to 3 degrees of freedom) system and the dynamical dependence on the initial mutual inclination between the orbital planes of the planets. In section 6, we discuss the limits of the applicability of our secular model and explore the neighborhood of the $\upsilon$ Andromedae system searching for the effects of mean-motion resonances in this region. We provide a summary of our results in section 7. Finally, in the Appendix we briefly describe the techniques used throughout  the  paper. 

\section{The 3--D model}

The Hamiltonian which describes the motion of the 3-body system in the astrocentric reference frame can be written, using Poincar\'e canonical variables, as
\begin{equation}
{\mathcal H}=\underbrace{\sum_{i=1}^2 (\frac{\vec{p}^{\,2}_i}{2\,m^\prime_i}-
\frac{\mu_i\,m_i^\prime}{|\vec{r}_i|})}_{{\rm Keplerian\, part}}-
\underbrace{\frac{G\,m_1\,m_2}{\Delta}}_{{\rm direct\,
part}}+\underbrace{\frac{(\vec{p}_1\cdot \vec{p}_2)}{m_0}}_{{\rm
indirect\, part}}, \label{eq:eq1}
\end{equation}
where $\vec{r}_i$ and $\vec{p}_i=m_i\,\frac{{\rm d}\vec{\rho}_i}{{\rm d}t}$ are the planet position vectors relative to the star and their conjugate momenta, respectively ($\vec{\rho}_i$ are the position vectors relative to the center of gravity of the three-body system). $G$ is the gravitational constant, $\mu_i=G(m_0+m_i)$, $m_i^\prime=m_i\,m_0/(m_0+m_i)$ and $\Delta = |\vec{r}_1-\vec{r}_2|$. Hereafter, the indices $i=1,2$ stand for the inner and outer planets, respectively.

Associated with the Keplerian part of the Hamiltonian, a set of mass-weighted Delaunay's elliptic variables is introduced as
$$
\begin{array}{r@{=}lc@{=}l}
M_i & {\rm mean\,\,anomaly,} & L_i &m_i^\prime\sqrt{\mu_ia_i}{\rm ,}\\
\omega_i & {\rm argument\,\,of\,\,pericenter,} & G_i  &L_i\sqrt{1-e_i^2}{\rm
,}\\
\Omega_i& {\rm longitude\,\,of\,node,} & H_i &
L_i\sqrt{1-e_i^2}\,\cos\,I_i{\rm ,}
\end{array}
$$
where $a_i$, $e_i$ and $I_i$ are the canonical astrocentric semi-major axes, eccentricities and inclinations of the planets, respectively. The relationship between canonical and usual osculating heliocentric orbital elements is described in detail in Ferraz-Mello {\it et al.} (2004). In the canonical variables, the Hamiltonian (\ref{eq:eq1}) can be written as 
\begin{equation}
{\mathcal H}=-\sum_{i=1}^2 \frac{\mu_i^2\,m_i^{\prime\,3}}{2\,L_i^2} -
\frac{G\,m_1\,m_2}{a_2}\,
R(L_i,G_i,H_i,M_i,\omega_i,\Omega_i), \label{eq:eq2}
\end{equation}
where the sum describes Keplerian motions of the planets and $R$ is the disturbing function.

To study the secular behavior of the system, we perform numerically the averaging of the Hamiltonian (\ref{eq:eq2}) with respect to the mean anomalies of the planets. The description of the numerical averaging procedure can be found in (MM2004). The secular Hamiltonian is then defined by
\begin{equation}
{\overline {\mathcal H}_{\rm sec}} =
-\frac{1}{{(2\pi)^2}}\int_0^{2\pi}\int_0^{2\pi}
\frac{G\,m_1\,m_2}{a_2} \,
R(L_i,G_i,H_i,M_i,\omega_i,\Omega_i)\,dM_1dM_2, \label{eq:eq3}
\end{equation}
where the contribution of the Keplerian part is constant and therefore need not be considered.

After the elimination of the short periodic terms, the averaged Hamiltonian ${\overline {\mathcal H}_{\rm sec}}$ does not depend on $M_1$ and $M_2$; consequently, $L_1$ and $L_2$ (thus $a_1$ and $a_2$) are constant in time. It is worth emphasizing that, due to the scissors' kind of the averaging used, the semi-major axes are constant only up to the first order in the planetary masses; this consideration is important to bear in mind when the results of the model are compared to those obtained through purely numerical integrations.

The secular Hamiltonian ${\overline {\mathcal H}_{\rm sec}}$ given by Eq.~\ref{eq:eq3} has four degrees of freedom and eight variables remain in the averaged problem. However, based on the conservation of the total angular momentum, we can reduce the problem by two degrees of freedom. This reduction is known as the elimination of nodes (Jacobi 1842).

For this purpose, we perform the canonical transformation of the nodes to
$$
\begin{array}{l@{=}l}
\theta_1 &\frac{\Omega_1+\Omega_2}{2}{\rm ,}\\
\theta_2 &\frac{\Omega_1-\Omega_2}{2}{\rm ,}\\
\end{array}
$$
and, accordingly,
$$
\begin{array}{l@{=}l@{=}l}
J_1&H_1+H_2&L_1\sqrt{1-e_1^2}\cos I_1+L_2\sqrt{1-e_2^2}\cos I_2{\rm ,}\\
J_2&H_1-H_2&L_1\sqrt{1-e_1^2}\cos I_1-L_2\sqrt{1-e_2^2}\cos I_2{\rm .}\\
\end{array}
$$
Since the averaged disturbing function $R$ depends on the nodes only through $\Delta\Omega=\Omega_1-\Omega_2$ (Brouwer and Clemence 1961), the angle $\theta_1$ is cyclic in the expression of $R$; hence, its conjugate action $J_1$ is a constant of motion.

In terms of canonical astrocentric elliptical elements, the components of the total angular momentum $\vec{C}$ are
$$
\begin{array}{r@{=}l}
C_x & \sum_{i=1}^2 {L_i\sqrt{1-e_i^2}\sin I_i\sin\Omega_i}{\rm ,}\\
C_y & \sum_{i=1}^2 {L_i\sqrt{1-e_i^2}\sin I_i\cos\Omega_i}{\rm ,}\\
C_z & \sum_{i=1}^2 {L_i\sqrt{1-e_i^2}\cos I_i}{\rm .}
\end{array}
$$
The direction of the constant total angular momentum defines a plane orthogonal to $\vec{C}$, which is an invariant of the problem (this plane is known as the Laplace invariable plane). The choice of the Laplace plane as the reference plane of coordinates implies that $C_x=C_y=0$ and $C_z=||\vec{C}||=C=const$.

If we assume that both planets orbit around the central star in the same sense, we obtain from the condition $C_x=C_y=0$ that the planetary eccentricities and inclinations are such that
\begin{equation}
G_1\sin I_1=G_2 \sin I_2. 
\label{eq:eq4}\end{equation}
The consequence of this condition is that, for $C_z > 0$, the variation of the inclinations of the planets is confined between zero and $90^\circ$, with a degeneracy of the problem at $I_i=90^\circ$.  The longitudes of the ascending nodes referred to the invariable plane are coupled in the following way:
$$
\Omega_1=\Omega_2\pm 180^\circ{\rm .}
$$
Two immediate consequences of this condition are 1) the system is described by the anti-aligned orientation of the nodal lines and 2) the problem is invariant under rotation of the angle $\Delta\Omega$ by $180^\circ$.

Based on the invariance of the direction of $\vec{C}$, we can partially reduce the system by one degree of freedom (Malige {\it et al}. 2002). In addition, using the invariance of the norm of the total angular momentum, we can perform the reduction by one more degree of freedom. Indeed, the condition given by $C=const$ constrains the action variables of the problem, in such a way that
\begin{equation}
\begin{array}{l@{=}l@{=}l}
J_1& H_1+H_2 & C{\rm ,}\\

J_2& H_1-H_2 & (G_1^2-G_2^2)/C{\rm .}\\
\end{array}
\label{eq:eq5}\end{equation}
Introducing the explicit dependence on $C$ in the Hamiltonian (\ref{eq:eq3}), we perform the total reduction and obtain, as a result, a two-degrees-of-freedom model. We note that the secular behavior of the totally reduced problem parameterized  by $C$ is independent upon the mutual inclination between the two orbits, $I_{\rm mut}$, whose value is defined by the planetary eccentricities, for fixed masses and semi-major axes, as
$$
\cos{I_{\rm mut}}=\frac{C^2-G_1^2-G_2^2}{2G_1G_2}{\rm .}
$$

Instead of the total angular momentum $C$, the use of the Angular Momentum Deficit ($\rm AMD$) is sometimes more appropriated. This quantity is defined as (Poincar\'e 1897, Laskar 2000)
\begin{equation}
{\rm AMD}=L_1+L_2-C=\sum_{i=1}^2{L_i(1-\sqrt{1-e_i^2}\cos I_i})\\
\label{eq:eq3_1}\end{equation}
and has a minimum value (zero) for circular co-planar orbits and increases with the eccentricities and inclinations.

In this work, we use both total and partial reductions to study the 3--D dynamics of the planetary system. In the first case, we choose a pair of independent action variables and the corresponding two angles. The other pair of actions can be easily obtained using Eqs.\,(\ref{eq:eq5}), for a given value of $C$. Here, we opt for two different pairs of the action variables: one is ($G_1$,$G_2$) (that is, $e_1$ and $e_2$) and other is ($G_1$,$H_1$) (that is, $e_1$ and $I_1$). In the case of partial reduction, the system has three degrees of freedom and the set of the variables used consists of the eccentricities of the planetary orbits and their mutual inclination. The individual inclinations of the planetary orbits are obtained using Eq.~(\ref{eq:eq4}).

In the following sections, we will apply the model to a system with the dynamical characteristics of the system $\upsilon$ Andromedae, restricted to the planets {\bf c} and {\bf d}. We adopt the following parameters: the masses $M=1.3M_{\odot}$, $m_1\sin{i}=1.83M_{\rm Jup}$ and $m_2\sin{i}=3.97M_{\rm Jup}$, for the central star, inner  and outer planets, respectively; the semi-major axes, $a_1=0.83\,{\rm AU}$ and $a_2=2.56\,{\rm AU}$. The description of the secular behavior of the system is completed by the initial values of the planet  eccentricities, $e_1=0.252$ and $e_2=0.31$ at $\Delta\varpi=0$; formal uncertainties in the eccentricities are of the order of 20\% (Ford 2005). 

Since the observational data do not constrain the individual orbital inclinations, we assume that both planetary orbits are observed edge-on, so $\sin i=1$ for each planet, where $i$ is the orbital inclination to the plane of the sky. This system will be referred to, along this paper, as the ``edge-on outer $\upsilon$ Andromedae system". The spatial configuration of this system is represented by two coplanar planetary orbits placed in the invariable plane, whose inclination $i$ to the sky plane is $90^\circ$. In this case, the total Angular Momentum of the system is tangent to the sky and, up to second order in masses, its modulus can be calculated using the formula:
$$
C = \sum_{i=1}^2 {m_i\sqrt{G\,m_0\,a_i\,(1-e_i^2)}}{\rm .}
$$
For the fixed masses and semi-major axes of the edge-on outer $\upsilon$ Andromedae system, the Angular Momentum Deficit (\ref{eq:eq3_1}) is $3\times 10^{-3}$, in units of the solar mass, astronomical unit and year. It is interesting to note that the edge-on configuration is characterized by a minimal value of the total Angular Momentum. Its magnitude increases with the rate $1/\sin{i}$,  due to the increasing planetary masses. The ${\rm AMD}$ also grows in this case.

We can show that, up to second order in masses, the secular behavior of the systems with small $I_{\rm mut}$ is independent on the inclination of the invariable plane to the plane of the sky. This is a consequence of the fact that the main features of secular motion of planetary systems are independent on the individual values of the planetary masses, but only on their ratio. To show this, we first remind that the secular dynamics is defined by such global quantities as total energy and the total angular momentum. Then, we point out that, up to second order in the masses, the secular energy (Eq.\,\ref{eq:eq3}) can be divided by the factor $Gm_1m_2/a_2$, while the total Angular Momentum Deficit (Eq.\,\ref{eq:eq3_1}) can be divided by $m_2\sqrt{Ga_2}$. As a result, we have one problem which depends only on the mass ratio and the semi-major axis ratio.

In the case of coplanar orbits, the mass ratio may be determined from the observations, being not affected by the indetermination of the individual masses. As a consequence, the results of the edge-on outer $\upsilon$ Andromedae system obtained in the study of the coplanar dynamics (MM2004) for AMD $= 3\times 10^{-3}$ are valid for non edge-on systems with AMD $= 3\times 10^{-3}/ \sin i$. If $I_{\rm mut} \neq 0$, the determination of the mass ratio is affected by the unknown factor $\sin i_1 / \sin i_2$, where $i_1$ and $i_2$ are the individual inclinations to the sky plane of the inner and outer orbits, respectively. However, if $I_{\rm mut}$ is small, the factor $\sin i_1 / \sin i_2$ can still be approximated by 1 that allows us to extend the results in the same way as described above.

Finally, throughout the paper, we will compare the results derived from our semi-analytical approach with direct numerical orbit integrations.

\section{Geometrical pictures of the 3--D secular dynamics}

In this section, we present the topology of the phase space of the Hamiltonian system given by Eq.~(\ref{eq:eq3}) and analyze its relationship with planetary dynamic.

First, to visualize the dynamical features of the fully-reduced system, we introduce a representative plane of initial conditions. The space of initial conditions of the two-degrees-of-freedom Hamiltonian system given by Eqs.~(\ref{eq:eq2})--(\ref{eq:eq3}) is four-dimensional, but the problem can be reduced to the systematic study of initial conditions in which two of them are kept constant. This plane may be chosen in such a way that all possible configurations of the system are included and thus all possible regimes of motion of the system under study can be represented on it.

We define the angular variables of the representative plane in the following way. The first of them comes from the studies of the planar secular problem (MM2004): it is
$\Delta\varpi =\varpi_1 -\varpi_2=\omega_1 -\omega_2 +\Delta\Omega$,
where $\varpi_1$ and $\varpi_2$ are the longitudes of the planetary pericenters and $\Delta\Omega =\pm 180^\circ$.  We know that the angle $\Delta\varpi$ can either circulate or oscillate about $0$ or $180^\circ$. In both cases, it goes through either $0$ or $180^\circ$ for all initial conditions. Hence, without loss of generality, the angular variable $\Delta\varpi$ can initially be fixed at $0$ or $180^\circ$.

The choice of the second angular variable, $2\,\omega_1$, is based on the particular property of the planetary disturbing function that requires the same parity of the indices preceding the planet arguments, $\omega_1$ and $\omega_2$ (Brumberg 1995). Indeed, using this property, we can re-write a generic periodic argument of the Hamiltonian function
(\ref{eq:eq3}), $k\,\omega_1 + l\,\omega_2 + m\,\Delta\Omega$,
as
$$\frac{k+l}{2}\,(2\,\omega_1) - l\,\Delta\varpi + (m+l)\,\Delta\Omega ,$$ where $\Delta\Omega =\pm 180^\circ$ and $k$, $l$ and $m$ are integers which vary within the interval ($-\infty$, $+\infty$). Our experimental studies have shown that the angular variable $2\,\omega_1$ can also either circulate or oscillate about $180^\circ$, which means that the argument of pericenter of the inner planet, $\omega_1$, either circulates or oscillates about $\pm 90^\circ$ (the sign depends on the chosen initial values of angle variables). In this case, it always goes through either $0$ or $\pm 90^\circ$ for all initial conditions. Hence, the initial values of $2\omega_1$ can be also fixed at $0$ or $180^\circ$.

Now, the dynamics of the twice-reduced Hamiltonian (\ref{eq:eq3}) can be represented on the plane of the initial values of the pair ($e_1\cos{\Delta\varpi}$,$e_2\cos{2\,\omega_1}$), where $e_1$ and $e_2$ are the averaged planetary eccentricities and the angles $\Delta\varpi$ and $2\,\omega_1$ are restricted to take the values $0$ or $180^\circ$. This plane will be hereafter referred to as ($e_1$, $e_2$) representative plane. The information on the initial values of the angular variables is given by the coordinate signs: positive (negative) values on x-axis correspond to $\Delta\varpi =0$ ($180^\circ$), while positive (negative) values on y-axis correspond to $2\omega_1 =0$ ($180^\circ$).

\begin{figure}[ht]
\centerline{\hspace{-1cm}\epsfxsize=12cm\epsffile{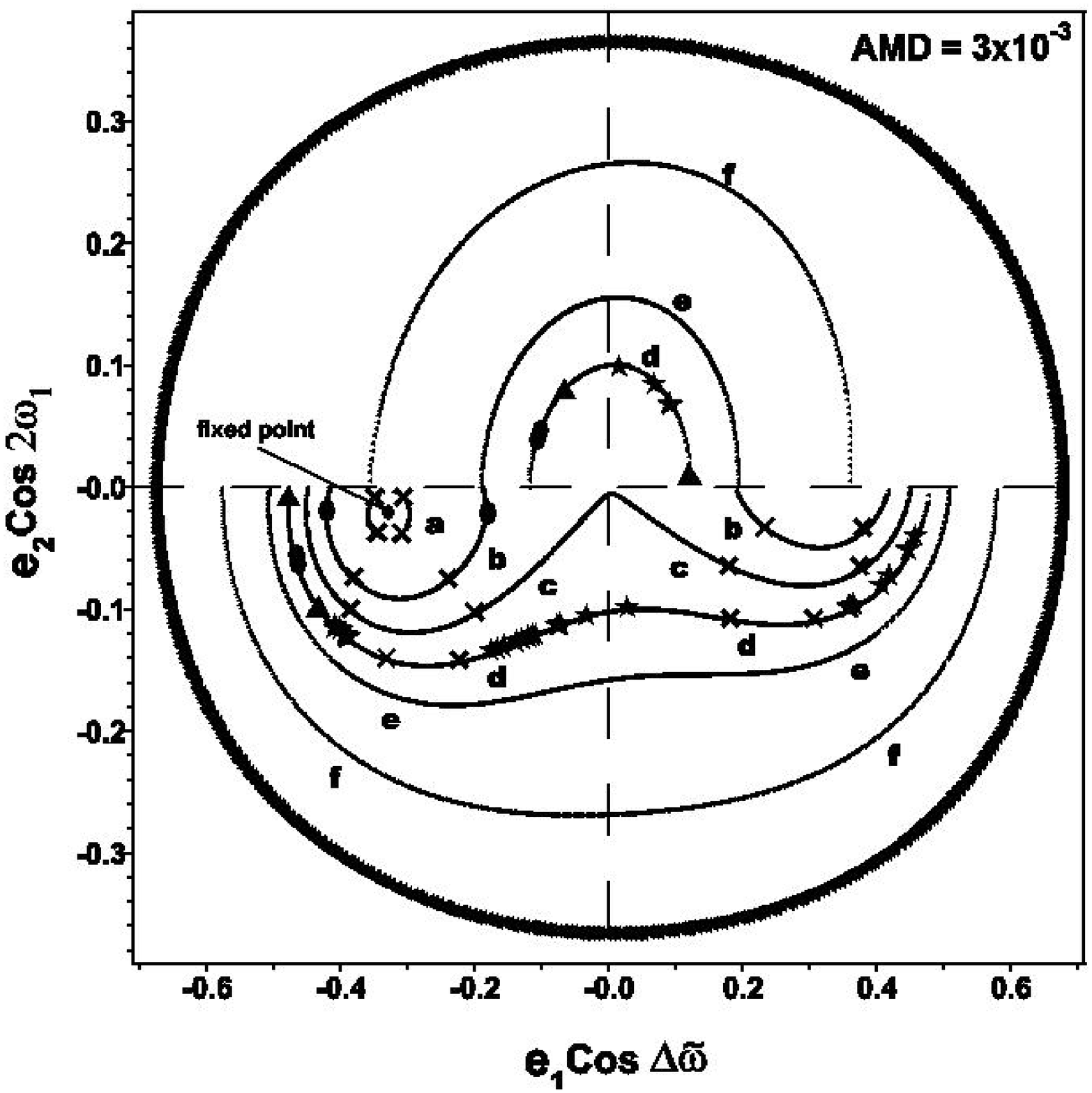}}
\caption{Michtchenko {\it et al.}}
\label{fig:fig1}
\end{figure}

Figure \ref{fig:fig1} shows the level curves of the energy given by the averaged secular Hamiltonian (\ref{eq:eq3}) on the ($e_1$,$e_2$)--plane. The level curves were calculated using ${\rm AMD}=3\times 10^{-3}$, which corresponds to the edge-on outer $\upsilon$ Andromedae system. The thick curve shows the boundary of the energy manifold defined by the chosen value of ${\rm AMD}$. The distinct spatial orientations of the system describing the lower and upper half-planes originate the discontinuity in the representative $(e_1,e_2)-$ plane, in the transition between these half-planes.

\mar{Fig.~1}

The geometrical analysis of the secular behavior is based on the fact that the secular energy ${\overline {\mathcal H}_{\rm sec}}$ and ${\rm AMD}$ are both conserved along one solution. Two other facts are also important: (1) Independently of whether the motions of $\Delta\varpi$ and $\omega_1$ are circulations or oscillations, all solution pass through the conditions $\sin\Delta\varpi=0$ ($\Delta\varpi=0$ and/or $180^\circ$) and $\sin 2\,\omega_1=0$ ($\omega_1=0$ and/or $90^\circ$); (2) As shown by numerical simulations, the variables $e_1$ and $e_2$ ($I_1$ and $I_2$) reach their maximal and minimal values at the condition $\cos\Delta\varpi = 0$ ($\cos 2\omega_1 =0$), allowing us to estimate the ranges of the eccentricity variation for both planets. As a consequence, an individual quasi-periodic solution intersects the representative plane at four points and all points must belong to one energy level. Of these points, one point is the chosen initial condition, and the other three point are its counterparts. The exceptional cases are: a) a stationary solution, which appears as a fixed point on the plane; b) periodic solutions, in which one of the angles is fixed, intersect the plane at only two points; and c) orbits of chaotic motion, which, due to diffusion processes, intersect an energy level on the ($e_1$,$e_2$)--plane at an arbitrary number of points.

The qualitative behavior of the angular variables, $\Delta\varpi$ and $\omega_1$, can be assessed from the geometrical analysis of the representative plane in Fig.~\ref{fig:fig1}. Concerning the behavior of the angle $\Delta\varpi$: if all intersections of an orbital path are located at the right-hand side of the ($e_1$,$e_2$)--plane, the angle $\Delta\varpi$ oscillates around zero. Conversely, if all points are located at the left-hand side of the plane, $\Delta\varpi$ oscillates around $180^\circ$. Finally, when intersections are found at both half--planes, the angle $\Delta\varpi$ is circulating.

In what concerns the behavior of the angle $\omega_1$, if all points of an orbital path are located at the lower half-plane, the angle $\omega_1$ oscillates around $\pm 90^\circ$. When the intersections are found in the lower and upper half-planes, the angle $\omega_1$ is circulating. The level of the maximal secular energy, ${\overline {\mathcal H}_{\rm sec}}=-1.02370\times 10^{-4}$, appears in Fig.~\ref{fig:fig1} as a fixed point with coordinates $e_1=-0.33$ and $e_2=-0.02$, corresponding to $I_{\rm mut}=42^\circ .9$. This is a stationary solution of the secular Hamiltonian, which is characterized by constant eccentricities and inclinations. In the close vicinity of the stationary solution, the angles $\Delta\varpi$ and $\omega_1$ (consequently, $\omega_2$) are oscillating around $180^\circ$ and $\pm 90^\circ$, respectively.

We show in Fig.~\ref{fig:fig1} six levels with decreasing energy denoted by letters from {\bf a} to {\bf f}. Several examples of intersections of planetary paths with the ($e_1$,$e_2$)--plane are shown by different symbols. The level {\bf a} (${\overline {\mathcal H}_{\rm sec}}=-1.02377\times 10^{-4}$), close to the stationary solution, is located at the quadrant of the plane corresponding to conditions $\Delta\varpi=2\omega_1=180^\circ$. The possible solutions in this case are oscillations of both angles $\Delta\varpi$ and $\omega_1$ (and $\omega_2$) around $180^\circ$ and $\pm 90^0$, respectively. Four crosses along the level show the intersection of one solution with the representative plane.

Along the level {\bf b}, with ${\overline {\mathcal H}_{\rm sec}}=-1.0247\times 10^{-4}$, the energy level splits into two unconnected branches: one is in the lower-left quadrant, while the other is in the lower-right quadrant; both are far away from the origin of the plane. The location of this level on the lower half-plane indicates that all solutions have $\omega_1$ oscillating around $\pm 90^\circ$. On the other hand, the secular angle $\Delta\varpi$ is allowed now to circulate or oscillate around $180^\circ$, depending on the initial conditions. One solution on this level, presented by four crosses, has the angle $\Delta\varpi$ in retrograde circulation. Initial conditions leading to solutions with $\Delta\varpi$ oscillating around $180^\circ$ with a very small amplitude are shown by full circles. As we will see later, the passage from oscillation to circulation of $\Delta\varpi$ in this case is merely kinematical and there is no real separatrix between two modes of motion, which evolve continuously from one to another. To make this fact clear in what follows, we will use the composite word ``circulation/oscillation" to indicate the regime of motion where the two kinds of motion co-exist, but are not topologically separated.

Along the level {\bf c}, with ${\overline {\mathcal H}_{\rm sec}}=-1.0257\times 10^{-4}$, most initial conditions are still characterized by the libration of $\omega_1$ around $\pm 90^\circ$ and circulation/oscillation of $\Delta\varpi$ (one example of this kind of motion is given by four crosses). The exceptions are solutions in the close vicinity of the origin of the plane, where the transition between the oscillating and circulating  modes of motion of $\omega_1$ occurs. To better illustrate this feature, we show in Fig.~\ref{fig:fig2} the surface of section and dynamic spectrum constructed along the energy level {\bf c}. The phase space (top panel) is dominated by the regime of $\omega_1$-libration, with the fixed point at $e_1\cos{2\omega_1}=-0.29$, corresponding to $e_2\cos{2\omega_1}=-0.1$ and $I_{\rm mut}=41^\circ .8$. A separatrix between circulation and libration of $\omega_1$ appears when $e_1\cos{2\omega_1}\sim -0.1$  in Fig.~\ref{fig:fig2} and a new regime of motion appears near the origin: this new regime correspond to a true secular resonance of the angle $\Delta\varpi$, which librates around $180^\circ$. The dynamic spectrum in Fig.~\ref{fig:fig2}\,{\it bottom} shows its structure with separatrix located at $e_1\cos{2\omega_1}=-0.1$ ($e_2\cos{2\omega_1}=-0.246$ and $I_{\rm mut}=34^\circ .7$).

\mar{Fig.~2}

This new dynamical feature is accentuated along the energy level {\bf d}, with ${\overline {\mathcal H}_{\rm sec}}=-1.027\times 10^{-4}$. The level is now a continuous curve on the lower half-plane, with a new branch appearing on the upper half-plane, where $\omega_1=0^\circ$. This geometry is also characteristic for the levels {\bf e}, with ${\overline {\mathcal H}_{\rm sec}}=-1.029\times 10^{-4}$, and {\bf f}, with ${\overline {\mathcal H}_{\rm sec}}=-1.038\times 10^{-4}$, in Fig.~\ref{fig:fig1}. Along the energy level {\bf d}, we have detected initial conditions leading to the solutions of three distinct regimes of motion, depending on the initial conditions. They are 1) a libration of $\omega_1$ around $\pm 90^\circ$ and a circulation of $\Delta\varpi$ (shown by crosses); 2) a libration of $\Delta\varpi$ around $180^\circ$ and a circulation of $\omega_1$ (full circles); 3) a circulation of both angles (triangles). The regions of transition between domains of different regimes are characterized by chaotic motion; one solution of chaotic motion is represented by star symbols randomly distributed along the energy level {\bf d}. To assess the type of motion of one solution, it is necessary to add an information obtained by direct numerical integrations of the equations of motion. It will be presented in the next section.

\section{Topology and dynamical maps of the totally reduced system: application to the edge-on outer $\upsilon$ Andromedae planets}

In this section we investigate the whole phase space of the Hamiltonian system given by Eq.\,\ref{eq:eq3}, with fixed values of ${\rm AMD}$. Our goal is to identify all possible regimes of its secular motion. For this task, we apply the model to one system whose masses and semi-major axes are those of the edge-on outer $\upsilon$ Andromedae system. The results are presented in the form of  topology and dynamical maps. The topology maps were calculated using the semi-analytical approach described in the previous sections, while the dynamical maps were obtained through numerical integrations of the exact equations of motion. 

To fully represent the secular dynamics, we have chosen two subspaces of initial conditions: one is the ($e_1$,$e_2$)--plane introduced in the previous section, and the other is a similar plane on which $e_2$ is substituted by the inclination of the inner planet over the invariable plane, the ($e_1$,$I_1$)--plane. In both cases, the initial values of the angular variables of the problem, $\Delta\varpi$ and $2\omega_1$, were fixed at either zero or $180^\circ$. The information on the initial values of the angular variables is given by the coordinate sign, as described in the previous section.

It is easy to show that both phase planes are equivalent: the solutions on one of these planes may be obtained from those on the other through Eqs.~(\ref{eq:eq5}), for a given ${\rm AMD}$. Nevertheless, we present two planes due to the fact that the ($e_1$,$e_2$)--plane is appropriate to display the details in high planetary inclinations, but the low inclination behavior is confined to the very narrow vicinity of the borders. At variance, the ($e_1$,$I_1$)--plane shows clearly the details of the low inclination dynamics.

\subsection {Phase portrait of the edge-on outer $\upsilon$ Andromedae system}

Figure \ref{fig:fig3} shows the topology ({\it left}) and dynamical maps ({\it right}) on the ($e_1$,$e_2$)--plane ({\it top}) and ($e_1$,$I_1$)--plane ({\it bottom}), both calculated for ${\rm AMD}=3\times 10^{-3}$, which corresponds to the edge-on outer $\upsilon$ Andromedae system. The location of the coplanar planets {\bf c} and {\bf d} is shown by a star symbol on all panels. The level curves of constant energy are plotted with solid lines in the topological maps. The mutual inclination between the initial orbital planes is given by dashed lines on the ($e_1$,$e_2$)--plane. On the ($e_1$,$I_1$)--plane, the dashed lines show the levels of $e_2=const$. For ${\rm AMD}=3\times 10^{-3}$, the system reaches the maximal value of the mutual inclination at $47^\circ .4$, when $e_1=e_2=0$, while the maximal value of the eccentricity of the outer planet is $0.365$, at $e_1=I_1=0$.

\mar{Fig.~3}

The main families of periodic orbits are plotted in the topological maps of Fig.~\ref{fig:fig3}\,({\it left}) by color dots: The red dots are associated with solutions of the Hamiltonian (\ref{eq:eq3}) in which the angle $\Delta\varpi$ is oscillating, while the blue dots to solutions with librating $\omega_1$. The stability of the periodic solutions was inferred from the features of the dynamical map (see description of the dynamical maps below). The character of motion on each orbit is assessed by the identification of the regimes of motion of the system by means of numerical integrations. Note that the oscillation of the secular angle $\Delta\varpi$ does not imply necessarily a resonant behavior, but may be just a kinematical continuation of the circulation regime of motion (see MM2004). It is worth mentioning that, except these circulation/oscillation solutions, the families present loci (stable by large dots and unstable by small dots) of the secular resonances in the phase space of the system and are in very good agreement with the location of the secular resonances obtained by numerical integrations. The calculation of the periodic solutions provides an important information about a dynamical system, without time-expensive numerical integrations. 

\mar{Fig.~4}

In Fig.~\ref{fig:fig3_1} we show four examples of periodic orbits: two orbits correspond to the red curves in Fig.~\ref{fig:fig3} and are, respectively, stable ({\it left-top}) and unstable ({\it left-bottom}). In the stable solution, the angle $\Delta\varpi$ librates around $180^\circ$. The other two orbits correspond to the blue curves and are, respectively, stable ({\it right-top}) and unstable ({\it right- bottom}). In the stable solution, the angle $\omega_1$ librates around $\pm 90^\circ$. 

In the construction of the dynamical maps of Fig.~\ref{fig:fig3}\,({\it right}), a grid of $72\times 79$ initial conditions was defined on the ($e_1$,$e_2$)--plane, with steps $\Delta e_1=0.02$ and  $\Delta e_2=0.01$. The initial inclinations were obtained from Eqs.~(\ref{eq:eq5}), for ${\rm AMD}=3\times 10^{-3}$. The initial value of the angular variables used were: from the set I in the right-top quadrant, the set II in the left-top quadrant, the set III in the left-bottom quadrant and the set IV in the right-bottom quadrant of the ($e_1$,$e_2$)--plane (see the definition of the sets in the Appendix). On the ($e_1$,$I_1$)--plane, a grid of $72\times 81$ initial eccentricity and inclination of the inner planet was defined with steps $\Delta e_1=0.02$ and $\Delta I_1=1^\circ$. The initial eccentricity and inclination of the outer planet were obtained using Eqs.~(\ref{eq:eq5}). The initial angular orbital elements of the planets are the same as used in the previous case.

The shading scale used in the dynamical maps in Fig.~\ref{fig:fig3}\,({\it right}) is related to the degree of stochasticity of the solutions: the lighter regions in the dynamical maps correspond to initial conditions of regular motion, darker tones indicate increasingly chaotic motion. The limit of the domains of allowed and forbidden motion can be calculated using Eqs.~(\ref{eq:eq5}): with conditions $I_1=I_2=0$ on the ($e_1$,$e_2$)--plane and $e_2=I_2=0$ on the ($e_1$,$I_1$)--plane. Outside, there are no solutions of the Hamiltonian system (\ref{eq:eq3}), for ${\rm AMD}=3\times 10^{-3}$. 

The analysis of the structure of the dynamical maps reveals interesting topological properties and reflect important features of the 3--D secular dynamics, when the orbital elements of the edge-on outer ${\upsilon}$ Andromedae system are used. The ($e_1$,$e_2$)-- and ($e_1$,$I_1$)--planes are dominated by a gray background of regular secular motion of the system. The narrow white strips coincide with the location of the stable periodic solutions of the secular system. The domains of chaotic motion appearing as black regions in Fig.~\ref{fig:fig3}\,({\it right}) are associated with separatrices of different regimes of secular motion (and unstable periodic orbits).

The careful analysis of the results of the numerical simulations, always accompanied by the study of the geometry of the energy levels and periodic solutions, allowed us to identify various different regimes of motion of the system. Their domains are marked by {\bf 1}--{\bf 4} on the dynamical maps in Fig.~\ref{fig:fig3}\,({\it right}). The domains {\bf 1} are regions of motion characterized by the coupled variation of the eccentricity and inclination of the inner planet and the libration of the angle $\omega_1$ around $\pm 90^\circ$. They are located on the lower half-planes at high mutual inclination (above $35^\circ$). The white strips follow the location of the periodic resonant solution in these regions. This regime of motion is often referred to as Kozai resonance in the literature. In this work, we designate this resonance as $e$--$I$ coupling, or Lidov--Kozai resonance, to pay homage to the Russian scientist, who first discovered this dynamical phenomenon (Lidov 1961). The stationary solution of the secular Hamiltonian (\ref{eq:eq3}), together with the paths of energy levels from {\bf a} to {\bf c} shown in Fig.~\ref{fig:fig1}, belong to the domains of this resonance. Inside the Lidov--Kozai resonance, the angle $\Delta\varpi$ shows either a retrograde circulation or an oscillation around $180^\circ$. Two examples of the periodic orbits (stable and close-to-unstable) in this regime of motion were shown in Fig.~\ref{fig:fig3_1}\,({\it right }). Note that the unstable orbit escapes from the Lidov-Kozai resonance after a while.

The domains of the Lidov--Kozai resonance are delimited in the phase space by regions of strong chaotic motion, which appear as black zones in the dynamical maps. In their close vicinity, at lower mutual inclinations, a new resonant regime of motion appears; its domains are labeled by {\bf 2} in Fig.~\ref{fig:fig3}. This resonance is characterized by the true libration of the secular angle $\Delta\varpi$ around $180^\circ$ and the prograde circulation of both angles $\omega_1$ and $\omega_2$. The family of the stable periodic solution of this resonance can be observed as the narrow white strips inside the domains {\bf 2}. Two examples of the periodic orbits (stable and close-to-unstable) in this regime of motion were shown in Fig.~\ref{fig:fig3_1}\,({\it left column}).

The regime of motion in the region {\bf 2} is better illustrated on a surface of section. In Fig.~\ref{fig:fig4}, we show the surface of section and dynamic spectrum constructed along the energy level ${\overline {\mathcal H}_{\rm sec}}=-1.033\times 10^{-4}$. The dominating regime of motion at this energy is the secular resonance of $\Delta\varpi$ described above, with the stable fixed point at $e_1\cos\Delta\varpi=-0.385$ ($e_2\cos{2\omega_1}=-0.187$ and $I_{\rm mut}=32^\circ .7$). The unstable fixed point of this resonance is located at $e_1\cos\Delta\varpi=0.433$ ($e_2\cos{2\omega_1}=-0.147$ and the same mutual inclination).

\mar{Fig.~5}

We can see in Fig.~\ref{fig:fig4} that the separatrix involving the secular resonance divides the whole domain of motion into two zones: the inner zone, close to the origin, and an outer zone. The inner zone shows a complex dynamical structure characteristic of the presence of secondary resonances. The dynamic spectrum of the solutions with initial conditions $e_1\sin\Delta\varpi=0$ (Fig.~\ref{fig:fig4}\,{\it bottom}) reveals the existence of islands of regular motion inside the sea of chaos in this region. The secular angles inside the inner zone are in circulation: retrograde for $\Delta\varpi$ and prograde for $\omega_1$ and $\omega_2$. The domains of this regime of motion are marked with the  label {\bf 3} in the dynamical maps of Fig.~\ref{fig:fig3}.

In the outer zone, the circulation of $\Delta\varpi$ inverts its direction. This regime of motion is characterized  by prograde circulation of all angles: $\Delta\varpi$, $\omega_1$ and $\omega_2$. The domains of this regime of motion are marked  with the label {\bf 4} in Fig.~\ref{fig:fig3}. This regime of motion extends to mutual inclinations close to $0$ and is known from the study of dynamics of the planar planetary model (MM2004). For some initial conditions, the angle $\Delta\varpi$ oscillates around $0$ or $180^\circ$; the regions of oscillation of $\Delta\varpi$ in the domains {\bf 4} are seen as the white strips on the dynamical maps in Fig.~\ref{fig:fig3}. It should be emphasized that the difference between circulation and oscillation of $\Delta\varpi$ in this region is merely kinematical and there are no true separatrices between the two modes of motion (MM2004). In other words, the apsidal alignment (or anti-alignment) in this case is a simple circulation around a center displaced from the origin and not a resonant motion. The 3--D dynamics of the system in the domains {\bf 4} is regular and nearly similar to that of the coplanar case, even for mutual inclinations as large as $30^\circ$. However, darker tones on dynamical maps in Fig.~\ref{fig:fig3}, also appear indicating increasingly chaotic motion. The explanation for this feature is the proximity of the system to mean-motion resonances of high order (see Section 7).

\mar{Table~I}

Finally, the regimes of motion appearing in Fig. 3 are summarized in Table I (the regime {\bf 5} will be discussed in the next section).

\subsection{Dependence on the Total Angular Momentum Deficit}

The secular behavior of the fully-reduced system is determined by the adopted value of the Total Angular Momentum Deficit. In the previous section we have shown the main features of the dynamics obtained for ${\rm AMD}=3\times 10^{-3}$, which corresponds to the edge-on outer $\upsilon$ Andromedae system. However, the exact value of this quantity is unknown, mainly, due to the fact that the observational data contain no information on the mutual inclination between the planets and the inclination of the invariable plane. 
For this reason, in this section, we investigate the evolution of the secular dynamics as a function of ${\rm AMD}$. 

We have chosen two distinct values: one is smaller (${\rm AMD}=1\times 10^{-3}$), and other is larger (${\rm AMD}=8\times 10^{-3}$), than AMD of the edge-on $\upsilon$ Andromedae system, notwithstanding that AMD of the actual outer $\upsilon$ Andromedae system cannot be smaller than the edge-on value. The smaller value of ${\rm AMD}$ is however considered in this work to complete the dynamical picture of the secular system. At variance, larger values of ${\rm AMD}$ are not impossible. For instance, it is enough to assume that both orbits are edge-on (to keep the same masses), but have a non-zero mutual inclination. (The chosen value is purposely large and corresponds to $I_{\rm mut}\approx 60^\circ$.) The main results obtained are given in the following.

For small values of ${\rm AMD}$, the dynamics of the system is very similar to that of the planar case. This is due to the fact that the values of the mutual inclination between the planetary orbits, allowed from Eqs.(\ref{eq:eq5}), in this case, are small. In Fig.~\ref{fig:fig5}, we present the topological picture of the phase space of the system with ${\rm AMD}=1\times 10^{-3}$, plotting the energy levels on the ($e_1$,$e_2$)-- and ($e_1$,$I_1$)--planes. The maximal value of the mutual inclination between orbital planes is $27^\circ$. The typical signatures of the high-inclination behavior, such as the existence of stationary solutions and bifurcations of the energy levels, shown in the previous sections, are not observed in Fig.~\ref{fig:fig5}. Only periodic solutions corresponding to the secular angle $\Delta\varpi$ exist.  These periodic solutions show the centers around which $\Delta\varpi$ oscillates (see MM2004).

\mar{Fig.~6}

At variance, for the large values of the Angular Momentum Deficit, the domain of possible motion of the system is extended to very high eccentricities and inclinations. Figure \ref{fig:fig6} shows the topology ({\it left}) and dynamical maps ({\it right}) on the ($e_1$,$e_2$)--plane ({\it top}) and ($e_1$,$I_1$)--plane ({\it bottom}), constructed for ${\rm AMD}=8\times 10^{-3}$. The notations used in this figure are analogous to those used in Fig.~\ref{fig:fig3}. The system reaches the maximal value of the mutual inclination at $79^\circ .7$, when $e_1=e_2=0$; the maximal value of the eccentricity of the outer planet is $0.58$, when $e_1=I_1=0$. The location of the fixed point (stationary solution) is at $e_1 =-0.81$ and $e_2=-0.1$, which correspond to $I_{\rm mut}=60^\circ .9$.

\mar{Fig.~7}

The geometry of the representative planes is similar to that obtained for ${\rm AMD}=3\times 10^{-3}$. The structure appearing in Fig.~\ref{fig:fig6}\,{\it left bottom}, in the low inclination region along the negative x-axis is an artifact. It only exists in the averaged model, disappearing when the full exact equations are considered.

Previous studies of the planar problem (MM2004) have shown that the large values of the Angular Momentum Deficit are characterized by the presence of a secular resonance with the angle $\Delta\varpi$ librating around $0$. Indeed, the domain of this resonance, denoted with the label {\bf 5}, is clearly seen on the dynamical map of the ($e_1$,$I_1$)--plane in Fig.~\ref{fig:fig6}\,{\it right bottom}; on the ($e_1$,$e_2$)--plane, this regime of motion occupies a narrow domain near the border.

The high-inclination regimes of motion {\bf 1} and {\bf 2}, discussed in the case of Fig.~\ref{fig:fig3}, are also present in Fig.~\ref{fig:fig6}. The regime {\bf 1}, that is, the Lidov--Kozai resonance, dominates  the lower half-planes. The location of the periodic orbits of this resonance is in good agreement with that calculated from our 3--D model (blue curves on the topological maps). The regime {\bf 2}, which is the high-inclination secular resonances of $\Delta\varpi$, dominates the upper half-planes: one in the right side, where $\Delta\varpi$ librates around $0$, and the other in the left side, where $\Delta\varpi$ librates around $180^\circ$.

\section{Partially reduced dynamical system: dependence on the mutual inclinations}

In the previous sections, the analysis of the 3--D dynamics was done with a system reduced to 2 degrees of freedom, whose evolution is constrained by a fixed value of the Angular Momentum Deficit. In this section, we consider the system only partially reduced to 3 degrees of freedom, where the direction of the angular momentum is taken into account, but not its modulus. In this case, we vary three orbital parameters of the problem; they are chosen as both planetary eccentricities and the mutual inclination of the planetary orbits. The study is done by means of dynamical maps and characteristic curves of periodic orbits of the Hamiltonian system given by Eq.~\ref{eq:eq3}.

Figures \ref{fig:fig7} and \ref{fig:fig8} show dynamical maps on the ($e_1$,$e_2$)--plane, obtained for three different values of the initial mutual inclination, namely $5^\circ$, $15^\circ$ and $45^\circ$. In the construction of the maps, a grid of $100\times 61$ initial eccentricities was defined on the ($e_1$,$e_2$)--plane. The angular variables used were: the set III, in the left-hand side of the panels, and the set IV, in the right-hand side of the ($e_1$,$e_2$)--plane (see the Appendix).

\mar{Fig.~8}

For $I_{\rm mut}=5^\circ$ and $15^\circ$, the domains of regular motion are characterized by the circulation/oscillation regime of motion of $\Delta\varpi$. The white regions of the phase space correspond to oscillation of $\Delta\varpi$: either around $0$ or around $180^\circ$. The regions coded in gray are regions of regular motion with circulating $\Delta\varpi$. Finally, the regions of highly nonharmonic and chaotic motion (dark zones) are domains of initial conditions leading to close approaches of the two planets.

\mar{Fig.~9}

The hatched regions are regions of large-scale instability followed by disruption of the system within the time-interval of 530,000 years. The location of the edge-on outer $\upsilon$ Andromedae system is shown by a star on both panels in Fig.~\ref{fig:fig7}. For these parameters, the system is within the domain where $\Delta\varpi$ oscillates around zero, as has been noted in previous studies (Malhotra 2002, Chiang \& Murray 2002, Ford {\it et al}. 2005). The effect of the mean-motion resonances of high orders may be noted as a  weak nonharmonic gray feature in the close proximity of the system on both panels in Fig.~\ref{fig:fig7}.

The domain of the true secular resonance reported by MM2004 is seen at the high eccentricity region of regular motion near the right border. The resonant orbits are protected from close approaches by coupled variation of the planet eccentricities and libration of $\Delta\varpi$. It is interesting to observe that the orbit of the inner planet inside this resonance can reach eccentricities as high as 0.95, but the system of two planets continues to be stable (at least, over the time-interval of 530,000 years).

The differences between the dynamical features revealed by two planes in Fig.~\ref{fig:fig7} (obtained for $I_{\rm mut}=5^\circ$ and $15^\circ$) are negligible. We conclude that the dynamics of the 3--D system varies slowly in the low-inclination region and is similar to the dynamics of the planar system studied in detail in (MM2004). This conclusion is in agreement with the results obtained by Lee and Peale (2003) for hierarchical planetary systems.

Significant changes in the dynamics of the system take place for  initial mutual inclinations larger than $30^\circ$. The dynamical map obtained with $I_{\rm mut}=45^\circ$ is shown in Fig.~\ref{fig:fig8}. The Lidov--Kozai resonance is now dominating over the whole domain of regular motion. The analytically calculated characteristic curves of the periodic orbits of this resonance (in blue color) are in very good agreement with the features of the dynamical map (white strips).

The secular angle $\Delta\varpi$ circulates for almost all initial conditions, but, depending on the initial conditions, there are also oscillations around $180^\circ$. These oscillations, which are just a kinematical continuation of circulations, occur in the low $e_2$ region, where the red curve shows their location in Fig.~\ref{fig:fig8}. In the very high eccentricity region, the red curve indicates  a narrow domain of stable motion: this is the domain of the true secular resonance characterized by libration of the angle $\Delta\varpi$ around $180^\circ$.

Finally, domains of chaotic motion appear in Fig.~\ref{fig:fig8} as inclined black strips  of variable width; they are associated with the effects of the mean-motion resonances in the neighborhood of the system.

\section{Mean-motion resonances in the neighborhood of the ${\upsilon}$ Andromedae Planetary System}

The applicability of the model of secular  dynamics presented in this work is restricted to domains free from the influence of mean-motion resonances. The averaging procedure of Eq.\,\ref{eq:eq3} removes, together with short-period terms, all effects of mean-motion commensurabilities; as a consequence, the  semi-analytical approach can not provide any information about mean-motion resonances and dynamical instabilities associated to them. The validity of our secular model, may be analyzed from the dynamical maps obtained by numerical integration of the exact equations of the motion, where the effects of mean-motion resonances in the neighborhood of the system under study are always present. For this reason, we decided to investigate the neighborhood of the ${\upsilon}$ Andromedae system looking for the peculiar features of mean-motion resonances.  

Special caution  was taken with the $\sin i$ indetermination in the masses of the planets. At variance with the secular case, there are no evidences that the resonant behavior depends only on the mass ratio; therefore the dynamical structure of the phase space may be strongly sensible to the inclination of the orbits to the sky plane. For this reason, we use two sets of the planetary masses: The current one corresponding to the edge-on outer ${\upsilon}$ Andromedae system, with the masses $m_1=1.83M_{\rm Jup}$ and $m_2=3.97M_{\rm Jup}$, and a second set adopting $\sin i=0.5$, which corresponds to the inclination of $30^\circ$ on the sky plane, with the masses $m_1=2\times 1.83M_{\rm Jup}$ and $m_2=2\times 3.97M_{\rm Jup}$. (Note that the mass ratio is the same in both sets).

In both cases, the initial orbital elements in the simulations were uniformly distributed on the ($a_2$,$e_2$)--plane of the semi-major axis and eccentricity of the outer planet, within the domains $2.4$\thinspace AU$\leq a_2 \leq 2.7$\thinspace AU ($\Delta a_2=0.005$\thinspace AU) and $0\leq e_2\leq 0.6$ ($\Delta e_2=0.01$), respectively. The initial inclination of the outer planet to the invariable plane was fixed at $1^\circ$. The semi-major axis and eccentricity of the inner planet were fixed at their current values, $a_1=0.83\,{\rm AU}$ and $e_1=0.252$. We have chosen the set I of initial values of angular variables, and, to have the motion referenced to the invariable plane, we have calculated the inclination of the inner planet using Eq.~(\ref{eq:eq4}). 

Figure \ref{fig:fig9} shows the dynamical maps of the neighborhood of the current position of the outer planet {\bf d}: on the left panel, for the first mass set, and on the right panel for the second set. Once again, the shading scale was used to distinguish between the regions of regular and chaotic motion. In both panels in Fig.~\ref{fig:fig9}, the hatched regions are characterized by large-scale instabilities followed by disruption of the system within the time-interval of 530,000 years. The domains of chaotic motion are associated with mean-motion resonances. We note the dominating presence of the 5:1 resonance: this is the resonance of the lowest order ($4$) in this region. Weaker resonances of higher orders appear as black strips of variable width: they are the 11:2, 16:3 and 21:4 resonances of orders 9, 13 and 17, respectively. The location of the ${\upsilon}$ Andromedae system is marked by a star symbol on both panels. 

\mar{Fig.~10}

We note that the edge-on system shown in Fig.\,\ref{fig:fig9}\,{\it left} is far away from the region of the strong 5:1 resonance; this fact may validate the use of the secular model in the study of this system. On the other hand, it is close to the weak 16:3 mean-motion resonance. We believe that the situation is not dramatic in this case, although the resonant effects can be still seen on dynamical maps. Due to its high order, the 16:3 resonance is very narrow; moreover, due to constraints imposed by the conservation of the total angular momentum, this system may survive for a time comparable with the age of the star (Michtchenko and Ferraz-Mello 2001). 

The situation may be considered as dramatic in the case of the second set of the adopted planetary masses, corresponding to  a $30^\circ$--inclined invariable plane (Fig.\,\ref{fig:fig9}\,{\it right}). In this case, the region of large-scale instabilities increases and all mean-motion resonances are shifted  towards the larger semi-major axes. The system under study is located, now, very close to the strong 5:1 mean-motion resonance. No doubt that, in this case, the application of the secular model to $\upsilon$ Andromedae would be hazardous.

\section{Conclusions}

This work presents an extension of the planar secular semi-analytical model introduced in (MM2004)  to the study of the three-dimensional dynamics of planetary systems. The basic technique is the numerical averaging over short-periods of the mutual interaction of the two planets. We emphasize that, in the present work, we do not use expansions of the disturbing function in power series of the orbital elements nor in Fourier series of the angular variables, such as done  classically (e.g. Brouwer and Clemence 1961, Murray and Dermott 1999). This means that the secular motion of the planets is described very precisely, without any restriction about the magnitude of their eccentricities, inclinations and the mutual distance. The only condition for the applicability of the model is that the system must be located far from a strong mean-motion resonance.

Due to the invariance of the total angular momentum in the general three-body problem, the 3--D secular system is reducible to a two-of-degrees-of-freedom dynamical system (total or Jacobi reduction). The behavior of the totally reduced system is governed by the adopted value of the total Angular Momentum Deficit (${\rm AMD}$). Recall that the ${\rm AMD}$ is equal to zero for circular co-planar orbits and increases with increasing eccentricities and/or inclinations. We have shown that, up to second order in masses, the 3--D phase space structure of the secular system does not depend on the individual values of the planetary masses and semi-major axes, but only upon the ratios of these quantities. For a given ${\rm AMD}$, the mutual inclination between the planetary orbits is a function of these parameters and the planetary eccentricities. For small mutual inclinations, the secular dynamics is independent of the orbital inclinations to the plane of the sky. This allows us to overcome partially the problem of the mass indetermination. 

Using the non-linear secular 3--D model, we have constructed geometrical pictures of the secular phase space of the two-planet system in terms of eccentricities and inclinations. Owing to the symmetries in the secular Hamiltonian, the phase space structure can be visualized in representative planes of initial conditions, with the initial angular elements $\Delta\varpi$ and $2\omega_1$ fixed at either 0 or $180^\circ$. The analysis of the topology of the phase space of the Hamiltonian system given by Eq.~(\ref{eq:eq3}) allow us to estimate the range of the eccentricity/inclination variations without time-expensive numerical integrations of the equations of motion. 

We have investigated the whole phase space of the Hamiltonian system given by Eq.\,\ref{eq:eq3} looking for all possible regimes of secular motion. For this task, we applied our approach to the specific case of the system with a dynamical parameters of two outer planets, {\bf c} and {\bf d}, of the edge-on $\upsilon$ Andromedae. The qualitative study was always supplemented by direct numerical integrations over a wide range of initial conditions and the interpretation of the results was always founded on the analytical approach.

The topology of the phase space of the system was investigated by means of several techniques, nominally: energy level maps , characteristic curves of families of periodic orbits , surfaces of section, dynamic spectra and dynamical maps.

In the following we summarize the important features of the 3--D dynamics of the secular system.

1. The low-to-moderate eccentricity and mutual inclination regime of motion (domain {\bf 4} with $e_1 < 0.6$ and $I_{\rm mut} < 30^\circ$). This  is a general non-resonant regime, similar to that found in the planar case. The systems always exhibit two main modes of secular motion, characterized by circulation of $\Delta\varpi$ or its oscillation around 0 or $180^\circ$. There are no real separatrices between the two modes of motion, whose solutions  evolve continuously from one type to another; the differences between these solutions are merely kinematical. The apsidal alignment or anti-alignment in this case is a simple circulation around a center displaced from the origin. The arguments of pericenter are in regular direct circulation. The edge-on outer $\upsilon$ Andromedae system is in this regime of motion and its secular behavior is stable; weak instabilities seem to  occur only in the vicinity of the high-order 16:3 mean-motion resonance. 

2. The high eccentricity and low-to-moderate inclination regime (domain {\bf 5} with $e_1 > 0.6$ and $I_{\rm mut} < 10^\circ$) is characterized by large-scale instabilities, due to close approaches of the planets, followed by disruption of the system within a few thousands of years. The only surviving  solutions in this region are those inside the nonlinear secular resonance and are bounded by a zero-frequency separatrix. The secular angle $\Delta\varpi$ librates around $0$ and the variation of $e_1$ and $e_2$ is strongly coupled. This feature of the secular dynamics of two-planet systems is the same that appeared in the studies of the planar problem  (MM2004).

3. The high inclination regimes of motion (domains {\bf 1} -- {\bf 3} with $I_{\rm mut} > 30^\circ$): complex dynamical behavior with the presence of several regimes of resonant motion. The dominating behavior is the $e$--$I$ coupling, or the Lidov--Kozai resonance, characterized by the coupled variation of the eccentricity and inclination of the inner planet and the libration of the angle $\omega_1$ around $\pm 90^\circ$. At variance with the analogous phenomenon in restricted problems, the variation of the planet inclinations and eccentricities is constrained by the total angular momentum conservation.  The large eccentricity/inclination excursions induced by the Lidov--Kozai resonance in restricted problems can not occur in the planetary problem. A regime of motion with $\Delta\varpi$ in the secular resonance also exists in the high-inclination region. In this case, the secular angle $\Delta\varpi$ librates either around $0$ or $180^\circ$. This regime is new and has no relation with the true secular resonance found in the planar problem by Michtchenko and Malhotra (MM2004) 

Finally, the limits of applicability of the model were assessed by the construction of dynamical maps.  We remind that the validity of the results derived from the secular approach may be compromised by effects of mean-motion resonances. We detect the presence of the several mean-motion resonances in the neighborhood of the $\upsilon$ Andromedae system: one is the low-order 5:1 resonance, and the others are all of very-high order with no significant effects on the secular motion of the system. The system with the masses of the edge-on outer $\upsilon$ Andromedae system is located far away from the strong 5:1 resonance. 

\section{Appendix. Applied techniques}

In this section we describe the main numerical techniques used in the study of the 3--D dynamics of the planetary system.

\paragraph {Energy level maps.}
The topology of the phase space of the Hamiltonian system can be studied by plotting the energy level curves on representative planes of initial conditions. For this purpose, the equation ${\overline {\mathcal H}}_{\rm sec} - {\mathcal H}^* = 0$ is solved numerically, for a given value of the energy ${\mathcal H}^*$, using a numerical procedure for root finding. The secular Hamiltonian function ${\overline {\mathcal H}}_{\rm sec}$ is given by Eq.~(\ref{eq:eq3}). Applying the total reduction, we choose a pair of action variables and the other pair is obtained through Eqs.~(\ref{eq:eq5}). The pair of independent variables may be, for instance, the pair ($G_1$,$G_2$) (equivalently, $e_1$,$e_2$). In this case, for given ${\mathcal H}^*$ and $\rm AMD$, $e_1$ can be easily obtained as a function of $e_2$. Varying ${\mathcal H}^*$, we calculate a family of the energy levels and plot it on the representative ($e_1$,$e_2$)--plane.

\paragraph {Characteristic curves of families of the periodic orbits.}
The location of the zero precession rates of the angular variables were obtained through the conditions:
\begin{equation}
\begin{array}{lcl}
 \dot\omega_1=\frac{\partial {\overline {\mathcal H}_{\rm sec}}}{\partial G_1}=0 {\rm ,}&
&\dot\omega_2=\frac{\partial {\overline {\mathcal H}_{\rm sec}}}{\partial G_2}=0 {\rm .}\\
\end{array}
\label{eq:eq6}
\end{equation}

By definition, the difference of the planet periastron longitudes is $\Delta\varpi =\omega_1 -\omega_2 +\Delta\Omega$, where $\Delta\Omega =180^\circ$. Consequently, we obtain the condition $\Delta\dot\varpi =\dot{\omega}_1-\dot{\omega}_2=0$ for the zero precession rate of the secular angle $\Delta\varpi$. The derivatives can be computed numerically for any given point of the phase space using the second-order differentiation scheme (Press {\it at al.} 1986).

It should be emphasized that, for the system with two or more degrees of freedom, the conditions (\ref{eq:eq6}) do not define families of periodic orbits. In this work, we present the curves corresponding to $\dot{\omega}_1=0$ and $\Delta\dot\varpi=0$. They show the domains of possible libration or oscillation of the corresponding angles.

A two-degrees-of-freedom dynamical system is characterized by two fundamental frequencies, consequently, a general (quasi-periodic) solution of this system is a composition of two independent modes of motion. In this case, the condition of periodicity will be satisfied when the amplitude of one of two modes tends to zero. The precise location of the zero-amplitude periodic trajectories can be calculated numerically from the position of the fixed points on the surfaces of section.

\paragraph {Numerical integrations.}
In order to obtain planetary motions, the exact equations of planetary motion were numerically integrated in the framework of the three-body general problem. The accurate RA15 integrator (Everhart 1985) was used. To compare the results of numerical investigations with the results given by the developed model, we use the canonical astrocentric orbital elements of the planets, instead the usual osculating astrocentric elements. The transformation between these two sets can be found in detail in (Ferraz-Mello {\it et al}. 2004).

The initial angular orbital elements of the inner ($i=1$) and outer ($i=2$) planets used in the dynamical maps were:

Set I ($\cos\Delta\varpi =1$ and $\cos 2\omega_1 =-1$):
$$
\begin{array}{c@{=}rc@{=}r}
\omega_1& -90^\circ & M_1 & 0^\circ\\
\omega_2&  90^\circ & M_2 & 0^\circ
\end{array}
$$
Set II ($\cos\Delta\varpi =-1$ and $\cos 2\omega_1 =-1$):
$$
\begin{array}{c@{=}rc@{=}rc@{=}r}
\omega_1& 90^\circ & M_1 & 180^\circ\\
\omega_2& 90^\circ & M_2 &   0^\circ
\end{array}
$$
Set III ($\cos\Delta\varpi =1$ and $\cos 2\omega_1 =1$):
$$
\begin{array}{c@{=}rc@{=}rc@{=}r}
\omega_1& -180^\circ& M_1 & 0^\circ\\
\omega_2&    0^\circ& M_2 & 0^\circ
\end{array}
$$
Set IV ($\cos\Delta\varpi =-1$ and $\cos 2\omega_1 =1$):
$$
\begin{array}{c@{=}rc@{=}rc@{=}r}
\omega_1& 0^\circ & M_1 & 180^\circ\\
\omega_2& 0^\circ & M_2 &0^\circ
\end{array}
$$
For all sets, the initial value of $\Delta\Omega =\Omega_1 -\Omega_2$ was fixed at $180^\circ$.

In order to apply the theory developed in this paper to the outer $\upsilon$ Andromedae system, we have adopted the following parameters: the masses $m_{\rm star}=1.3M_{\rm Sun}$, $m_1=1.83M_{\rm Jup}$ and $m_2=3.97M_{\rm Jup}$; the semi-major axes $a_1=0.83\,{\rm AU}$ and $a_2=2.56\,{\rm AU}$; and the planet eccentricities $e_1=0.25$ and $e_2=0.34$ at $\Delta\varpi=0$ and $2\omega_1=0$. 

To remove the short-period oscillations (those of the order of the planetary orbital periods), a low-pass filtering procedure was implemented on-line with the numerical integration as described in detail by Michtchenko and Ferraz-Mello (1995). The numerical integrations were performed over a time interval of 524,544 years, which was large enough to allow an accurate and efficient averaging of the long-period effects, and also to detect the occurrence of secular resonances.

\paragraph {Dynamical maps.}
The orbits of the planets obtained trough direct numerical integrations were Fourier-transformed using the standard FFT algorithm. The information contained in the power spectra of the orbital elements was used in the construction of dynamical maps (see Michtchenko {\it et al}. 2002).

The mapped quantity is the spectral number $N$ defined as the number of peaks in the power spectrum of one calculated planetary orbit above an arbitrarily defined "noise". In this work, we consider as "noise" those peaks with amplitudes smaller than 5\% of the amplitude of the largest peak. The spectral number $N$ is used to qualify the chaoticity of planetary motion in the following way: small values of $N$ correspond to regular motion, while the large values indicate the onset of chaos.

Once the spectral numbers $N$ are determined for all the initial conditions on the grid, we plot them on the representative planes using a shading scale. The calculated values of $N$, in the range from 1 to 80, are coded by a gray level scale that varied linearly from white ($N=1$) to black ($N=80$). Since large values of $N$ indicate the onset of chaos, the shading scale is related to the degree of stochasticity of the initial conditions: lighter regions on the dynamical maps correspond to regular motion, darker tones indicate increasingly chaotic motion.

\paragraph {Surfaces of section and dynamic power spectra.}
The structure of the phase space of the two-degrees-of-freedom system was studied using the standard technique of the construction of surfaces of section.

Two sections have been chosen for presentation of the totally reduced system: The first is a section by the plane $\sin 2\omega_1=0$ and its coordinates are $e_1\cos{\Delta\varpi}$ and $e_1\sin{\Delta\varpi}$. The second is a section by the plane $\sin\Delta\varpi=0$ and its coordinates are $e_1\cos{2\omega_1}$ and $e_1\sin{2\omega_1}$.

It should  be emphasized that, in the construction of the surfaces of section, we have used as input the planetary solutions obtained by numerical integration  (the application of on-line filtering procedure is crucial in this task). For this reason, the erratic scatter of points due to the loss of numerical accuracy can be sometimes observed.

The dynamic power spectrum is a technique which is complementary to the surfaces of section and is very efficient to identify such phenomena as bifurcation and chaoticity. A dynamic spectrum presents, for each value of a given dynamical parameter (the abscissa of the plot), the evolution of the main oscillation modes of the planetary motion as a function of the parameter. Over the domains of regular motion, the proper frequencies vary continuously when the parameter is gradually varied. When the region of chaotic motion is approached, the frequency evolution shows a discontinuity, characterized by the erratic scatter of values.

\section*{Acknowledgments}
We thank Dr. Renu Malhotra for suggesting the problem and the analysis method. This work has been supported by the Brazilian National Research Council - CNPq, as well as the S\~ao Paulo State Science Foundation - FAPESP. The authors gratefully acknowledge the support of the Computation Center of the University of S\~ao Paulo (LCCA-USP) for the use of their facilities.

\vspace{-0.5cm}
\section*{\centering { \normalsize REFERENCES}}
\vspace{-0.2cm}

\begin{list}{}{ \setlength{\leftmargin}{0.5cm}
        \setlength{\itemindent}{-0.5cm}}

\item
Brouwer, D., and G.M. Clemence 1961. "Methods of Celestial Mechanics". Academic Press Inc., New York.

\item
Brumberg, V. A. 1995. "Analytical Techniques of Celestial Mechanics". Springer, Berlin.

\item
Butler, R.P., G.W. Marcy, E. Williams, H. Hauser, and P. Shirts 1997. Three New "51 Pegasi--Type" Planets. {\it ApJ} {\bf 474}, L115-L118.

\item
Butler, R.P., G.W. Marcy, D.A. Fischer, T.M. Brown, A.R. Contos, S.G. Korzennik, P. Nisenson, and R.W. Noyes 1999. Evidence for Multiple Companions to $\upsilon$Andromedae. {\it ApJ} {\bf 526}, 916-927.

\item
Chiang, E.I., S. Tabachnik, and S. Tremaine 2001. Apsidal Alignment in ${\upsilon}$  Andromedae. {\it AJ} {\bf 122}, 1607-1615.

\item
Everhart, E., 1985. An efficient integrator that uses Gauss-Radau spacings. \textit{Proceedings of IAU Colloq. 83} 115, 185-202.

\item
Ferraz-Mello, S., 1994. The convergence domain of the Laplacian expansion of the disturbing function. {\it Celest. Mech. Dynam. Astr.} {\bf 58}, 37-52.

\item
Ferraz-Mello, S., T.A. Michtchenko, and C. Beaug\'e  2004. Regular motions in extra-solar planetary systems. NATO ASI Series C: Mathematical and Physical Sciences. , to be published in "Chaotic Worlds: From Order to Disorder in Gravitational N-Body Systems" (B.A.Steves, ed.), Kluwer Acad. Publ. (in press).

\item
Ferraz-Mello, S., T.A. Michtchenko, and C. Beaug\'e  2005. The Orbits of the Extrasolar Planets HD 82943\,c and b. {\it ApJ} {\bf 621}, 473-481.

\item
Ford, E.B., B. Kozinsky, and F. Rasio 2000. Secular Evolution of Hierarchical Triple Star Systems. {\it ApJ} {\bf 535}, 385-401.

\item
Ford, E.B. 2005. Quantifying the uncertainty in the orbits of extra-solar planets. {\it AJ} {\bf 129}, 1706-1717.

\item
Ford, E.B., V. Lystad, and F.A. Rasio 2005. Planet-planet scattering in the upsilon Andromedae system. {\it Nature} {\bf 434}, 873-876.

\item
Harrington, R.S., 1969. The stellar three-body problem. {\it Celestial Mechanics} {\bf 1}, 200-209.

\item
Jacobi, C.G.J. 1842. ``Sur l'elimination des noeuds dans le probl\'eme des trois corps".  {\it Astronomische Nachrichten}, Bd XX , 81-102.

\item
Konacki, M., and A. Wolszczan 2003. Masses and orbital inclinations of planets in the PSR B1257+12 system. {\it ApJ} {\bf 591}, L147-L150.

\item
Laplace, P.S. 1799. ``M\'ecanique C\'eleste". English translation by N. Bowditch, Chelsea Pub. Comp. Edition, N.Y., 1966.

\item
Laskar, J. 1987. Secular evolution of the solar system over 10 millions years. {\it Astronomy and Astrophysics} {\bf 198}, 341-362.

\item
Laskar, J., and Robutel, P. 1995. Stability of the Planetary Three-Body Problem. I. Expansion of the Planetary Hamiltonian. {\it Celest. Mech. Dynam. Astr.} {\bf 62}, 193-217.

\item
Laskar, J. 2000. On the Spacing of Planetary Systems. Physical Review Letters {\bf84}, 3240-3243.

\item
Lee, M.H., and S.J. Peale 2003. Secular Evolution of Hierarchical Planetary Systems.  {\it ApJ} {\bf 592}, 1201-1216.

\item
Lidov, R. 1961. Analiz evolucii orbit iskustevennich sputnikov. {\it Problemi dvigenia iskustvennich nebesnich tel}, Izd. Akademii Nauk SSSR, Moscow (1963), 119-134.

\item
Lissauer, J. J. 1999. Three planets for Upsilon Andromedae. {\it Nature}  {\bf 398}, 659-660.

\item
Lissauer, J. J. and E.J. Rivera 2001. Stability Analysis of the Planetary System Orbiting $\upsilon$ Andromedae. II. Simulations Using New Lick Observatory Fits. {\it ApJ}  {\bf 554}, 1141-1150.

\item
Malhotra, R. 2002. A Dynamical Mechanism for Establishing Apsidal Resonance. {\it ApJ}  {\bf 575}, L33-L36.

\item
Malige, F., P. Robutel, and J. Laskar. 2002. Partial Reduction in the N-Body Planetary Problem using the Angular Momentum Integral. {\it Celest. Mech. Dynam. Astron.}  {\bf 84}, 283-316.

\item
Mazeh, T., Zucker, S., Torre, A. D., and van Leeuwen, F. 1999. Analysis of the {\it Hipparcos} measurements of $\upsilon$ Andromedae: a mass estimate of its outermost known planetary companion. {\it ApJ} {\bf 522}, L149-L151.
\item
Michtchenko, T.A., and S. Ferraz-Mello 1995. Comparative study of the asteroidal motion in the 3:2 and 2:1 resonances with Jupiter. I. Planar model. {\it Astronomy and Astrophysics} {\bf 303}, 945-963.

\item
Michtchenko, T.A., and S. Ferraz-Mello 2001. Resonant structure of the outer Solar System in the neighborhood of the planets. {\it AJ} {\bf 122}, 474-481.

\item
Michtchenko, T.A., D. Lazzaro, S. Ferraz-Mello, and F. Roig 2002. Origin of the basaltic asteroid 1459 Magnya: A dynamical and mineralogical study of the outer main belt. {\it Icarus} {\bf 158}, 343-359.

\item
Michtchenko, T.A., and R. Malhotra 2004. Secular dynamcs of the three-body problem: application to the $\upsilon$ Andromedae planetary system. {\it Icarus} {\bf 168}, 237-248.

\item
Poincar\'e, H. 1897. Sur une forme nouvelle des \'equations du probl\`eme des trois corps. {\it Bull.Astron.} {\bf 14}, 53-67.

\item
Press, W.H., Flannery, B.P., Teukolsky, S.A, and Vetterling, W.T. 1986. Numerical Recipes. Cambridge University Press.

\item
Stepinski, T. F., R. Malhotra, and D.C. Black 2000. The $\upsilon$ 
Andromedae System: Models and Stability. {\it ApJ} {\bf 545}, 1044-1057.

\item
Veras, D., and Armitage, P.J. 2004. The dynamics of two massive planets on inclined orbits. {\it Icarus} {\bf 172}, 349-371.

\end{list}

\newpage

\begin{table*}
\caption{The main regimes of secular motion. The behavior of the angles $\omega_1$ and $\Delta\varpi$ is identified by symbols: L -- libration, CP and CR -- prograde and retrograde circulation, respectively, O -- oscillation. }

\vspace{0.5cm}
\smallskip
\vbox{\tabskip=0pt
\offinterlineskip
\def\tableru{\noalign{\hrule height2pt}}
\def\tabler{\noalign{\hrule height1pt}}
\halign to 5.0 true in
{\strut#&
\tabskip=1em plus 2em&
#\hfil&
#\hfil&
#\hfil&
#\hfil&
\hfil#\hfil\cr
\tableru
\noalign{\smallskip}
&& Label & $\omega_1$ & $\Delta\varpi$ & $I_{\rm mut}$ & $e_1$ \cr
\tabler
\noalign{\smallskip}
&& {\bf 1} & L ($\pm 90^\circ$) & CR / O ($180^\circ$) & $> 35^\circ$ & all \cr
\tabler
\noalign{\smallskip}
&& {\bf 2} & CP & L ($0$ and $180^\circ$) & $> 30^\circ$ & all \cr
\tabler
\noalign{\smallskip}
&& {\bf 3} & CP & CR              & $\sim 30^\circ$ & all \cr
\tabler
\noalign{\smallskip}
&& {\bf 4} & CP & CP / O ($0$ and $180^\circ$) & $< 30^\circ$ & $< 0.6$ \cr
\tabler
\noalign{\smallskip}
&& {\bf 5} & CP & L ($0$) & $< 10^\circ$ & $> 0.6$ \cr
\noalign{\smallskip}
\tableru}}
\end{table*}

\clearpage

\section*{\centering {\normalsize Figure Captions}}

Figure 1. Schematic details of the phase space on the representative ($e_1$,$e_2$)-plane of initial conditions obtained for ${\rm AMD}=3\times 10^{-3}$. Positive (negative) values of $e_1$ correspond to $\Delta\varpi$ equal to $0$ ($180^\circ$); positive (negative) values of $e_2$ correspond to $2\omega_1$ equal to $0$ ($180^\circ$). The distinct spatial orientations of the system describing the lower and upper half-planes originate the discontinuity in the transition between these half-planes. Six levels, from {\bf a} to {\bf f}, with decreasing energy, are shown. The fixed point is a stationary solution of the Hamiltonian (\ref{eq:eq3}). The intersections of some numerically calculated orbits with the plane are shown by crosses, full circles, triangles and stars symbols (see text for further details).
\vspace{0.5cm}

\hspace{-0.5cm}Figure 2. ({\it Top:}) Surface of section defined by $\sin\Delta\varpi=0$ and ${\rm AMD}=3\times 10^{-3}$. ({\it Bottom}): Dynamic spectrum corresponding to the solutions whose initial conditions are on the axis $e_1=0$ of the surface of section.
\vspace{0.5cm}

\hspace{-0.5cm}Figure 3. {\it Left}: In black lines, energy levels of the secular Hamiltonian given by Eq.~(\ref{eq:eq3}) on the ($e_1$,$e_2$)--plane ({\it top}) and ($e_1$,$I_1$)--plane ({\it bottom}), for ${\rm AMD}=3\times 10^{-3}$. The signs $+$ or $-$, preceding the variable $e_1$, indicate that the initial values of $\Delta\varpi$ are zero or $180^\circ$, respectively. The signs $+$ or $-$, preceding the variables $e_2$ and $I_1$ indicate that the initial values of $2\omega_1$ are zero or $180^\circ$, respectively. In dashed lines, level curves of the mutual inclination on the ($e_1$,$e_2$)--plane and the eccentricity of the outer planet ($e_1$,$I_1$)--plane. Periodic solutions of the secular Hamiltonian, corresponding to oscillating $\Delta\varpi$ or librating $\omega_1$ are shown by red and blue dots (large for stable and small for unstable), respectively. The location of the edge-on outer $\upsilon$ Andromedae system is  indicated by a star. {\it Right}: Dynamical maps. The domains of the different regimes of motion are: {\bf 1} -- the Lidov--Kozai resonance, where the angle $\omega_1$ is in libration around $\pm 90^\circ$ and $\Delta\varpi$ is in retrograde circulation; {\bf 2} -- the secular resonance, where the angle $\Delta\varpi$ is in libration around $180^\circ$ and $\omega_1$ is in circulation; {\bf 3} -- the angles $\omega_1$ and $\Delta\varpi$ are in direct and retrograde circulation, respectively; {\bf 4} -- $\Delta\varpi$ is in the general regime of circulation/oscillation of the planar case and both arguments of pericenter are in direct circulation. The domains of forbidden motion, with no solutions of the Hamiltonian system for ${\rm AMD}=3\times 10^{-3}$, are hatched.
\vspace{0.5cm}

\hspace{-0.5cm}Figure 4.  Four of the periodic solutions shown in Fig.~\ref{fig:fig3}. Each panel presents two curves: thick curve is defined on the ($e_1\cos{\Delta\varpi}\times e_1\sin{\Delta\varpi}$)--plane and thin on the ($I_1\cos{2\omega_1}\times I_1\sin{2\omega_1}$)--plane. {\it Left column:} The stable ({\it top}) and unstable ({\it bottom}) orbits with the secular angle $\Delta\varpi$ librating around $180^\circ$ and/or circulating. {\it Right column:} The stable ({\it top}) and unstable ({\it bottom}) orbits with the angle $\omega_1$ librating around $\pm 90^\circ$ and/or circulating. 
\vspace{0.5cm}

\begin{figure}[ht]
\centerline{\hspace{-1cm}\epsfxsize=8cm\epsffile{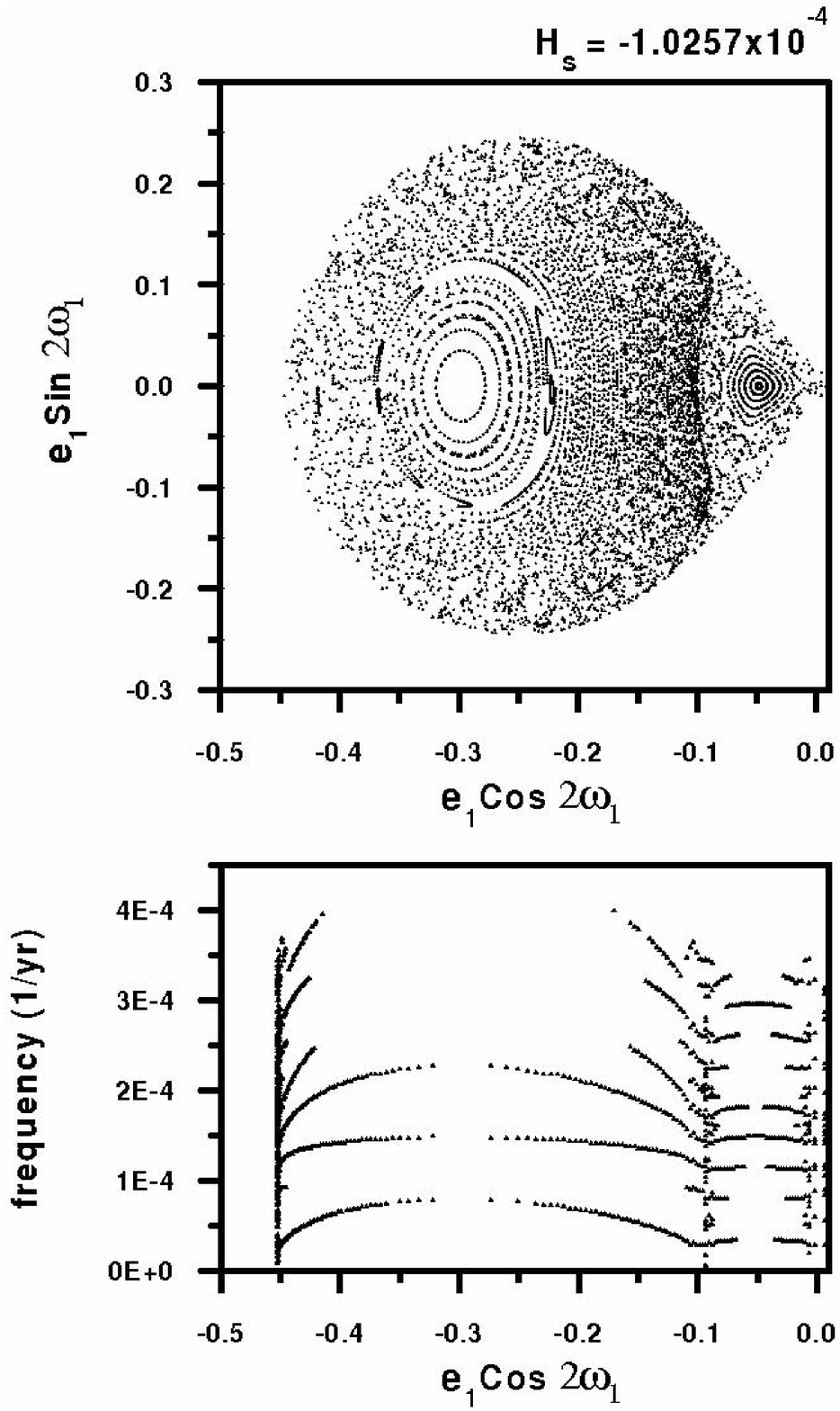}}
\caption{Michtchenko {\it et al.}}
\label{fig:fig2}
\end{figure}

\begin{figure}[ht]
\centerline{\hspace{-1cm}\epsfxsize=15cm\epsffile{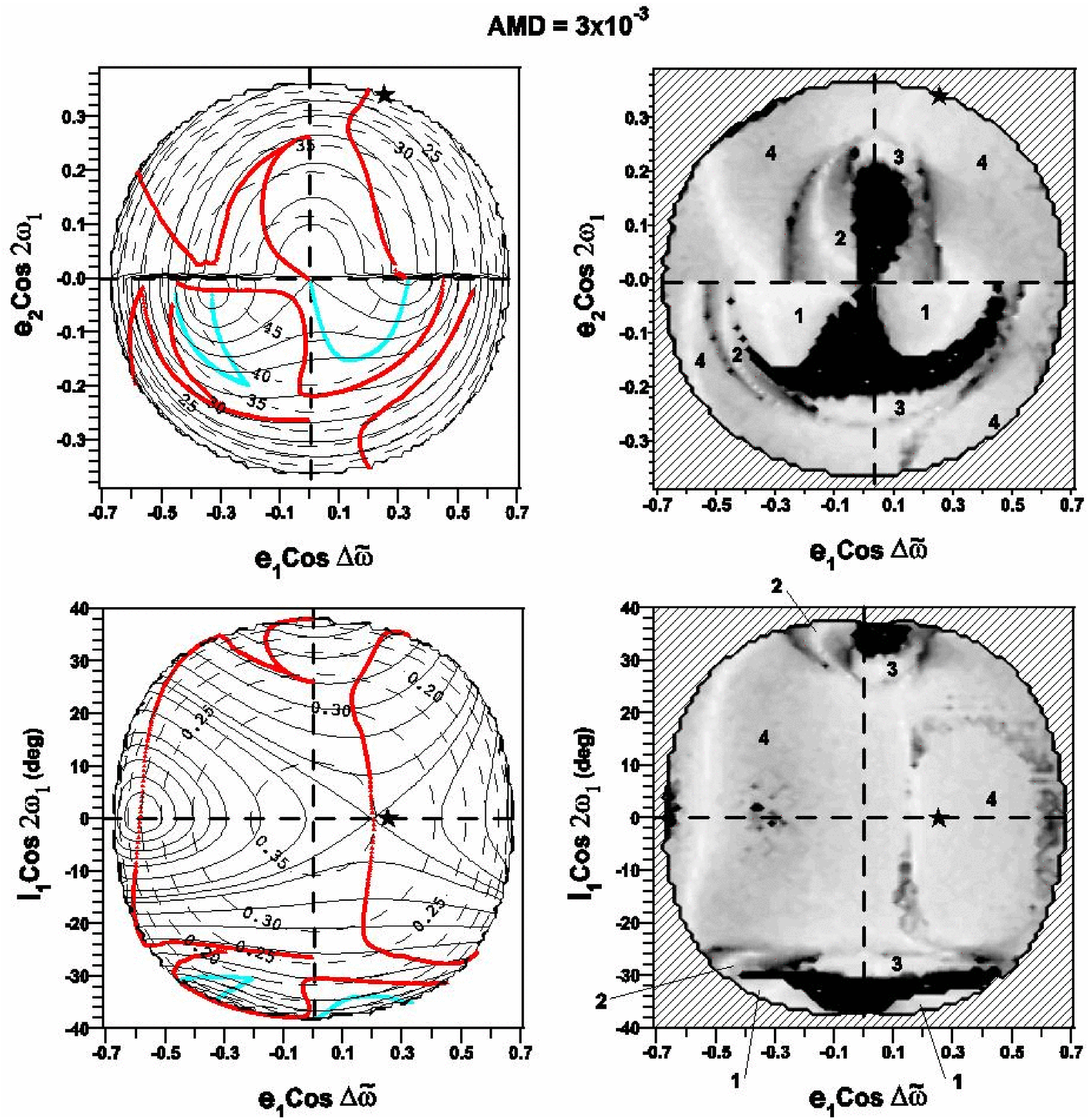}}
\caption{Michtchenko {\it et al.}}
\label{fig:fig3}
\end{figure}

\begin{figure}[ht]
\centerline{\hspace{-1cm}\epsfxsize=15cm\epsffile{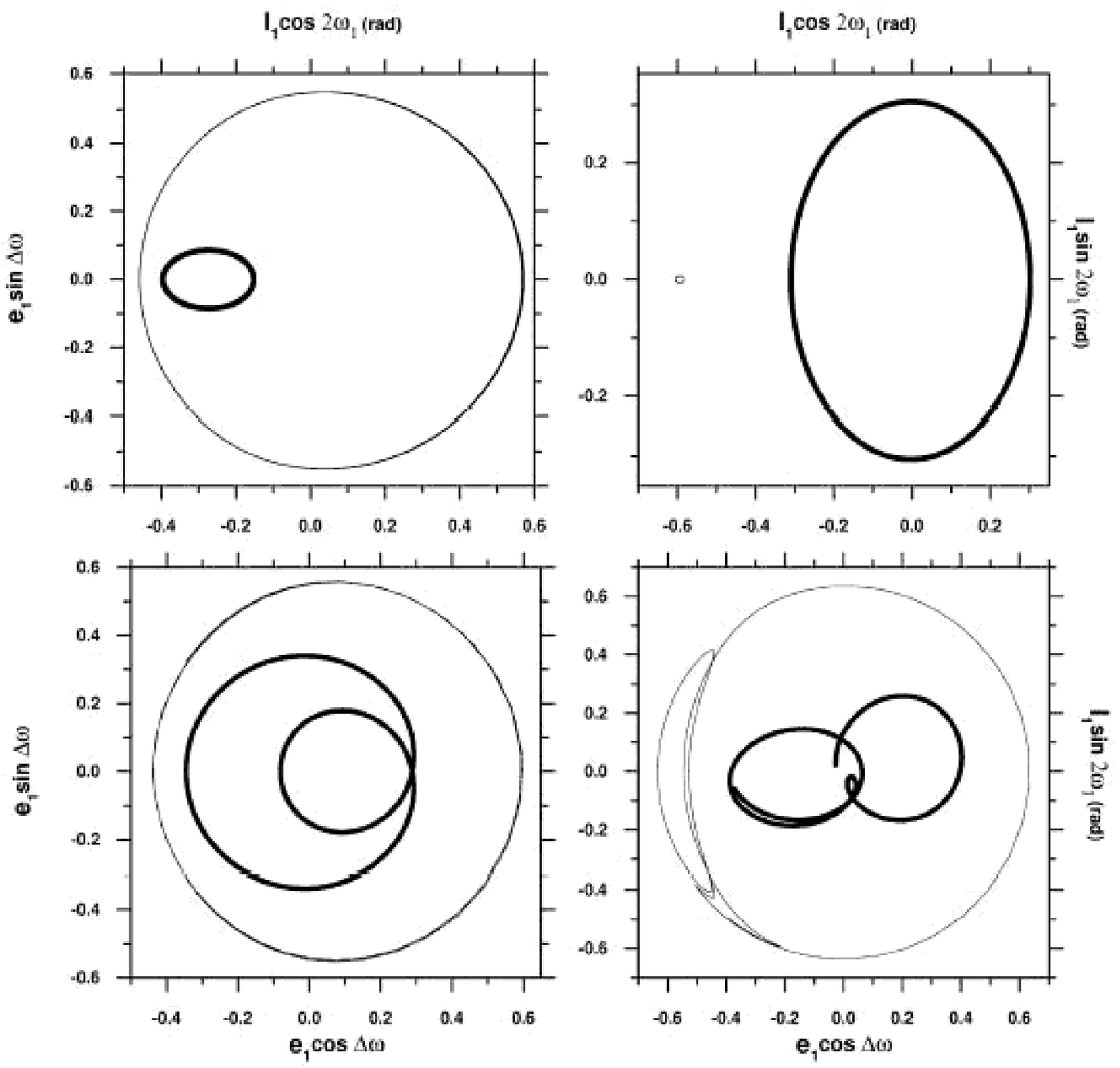}}
\caption{Michtchenko {\it et al.}}
\label{fig:fig3_1}
\end{figure}

\begin{figure}[ht]
\centerline{\hspace{-1cm}\epsfxsize=8cm\epsffile{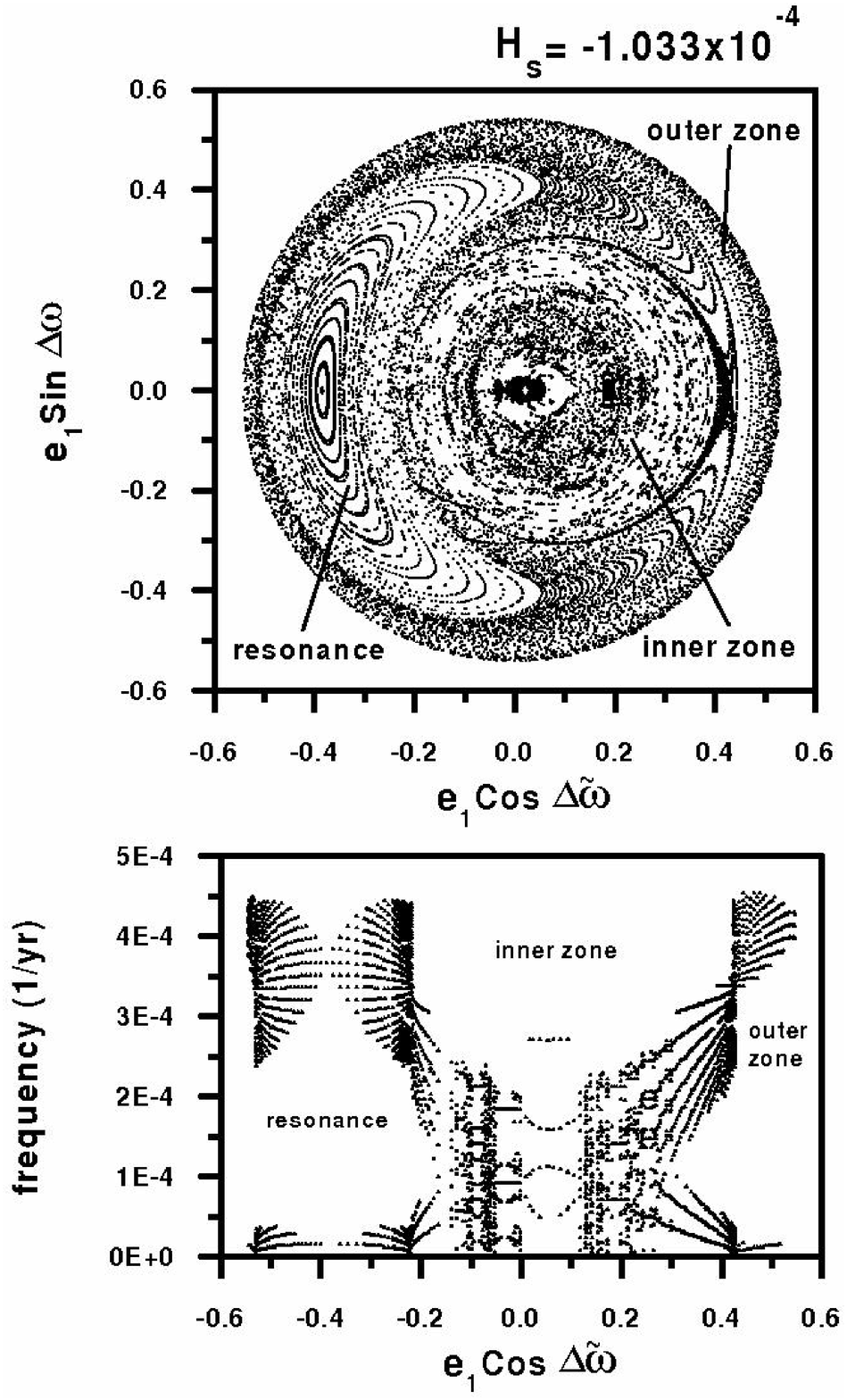}}
\caption{Michtchenko {\it et al.}}
\label{fig:fig4}
\end{figure}

\begin{figure}[ht]
\centerline{\hspace{-1cm}\epsfxsize=8cm\epsffile{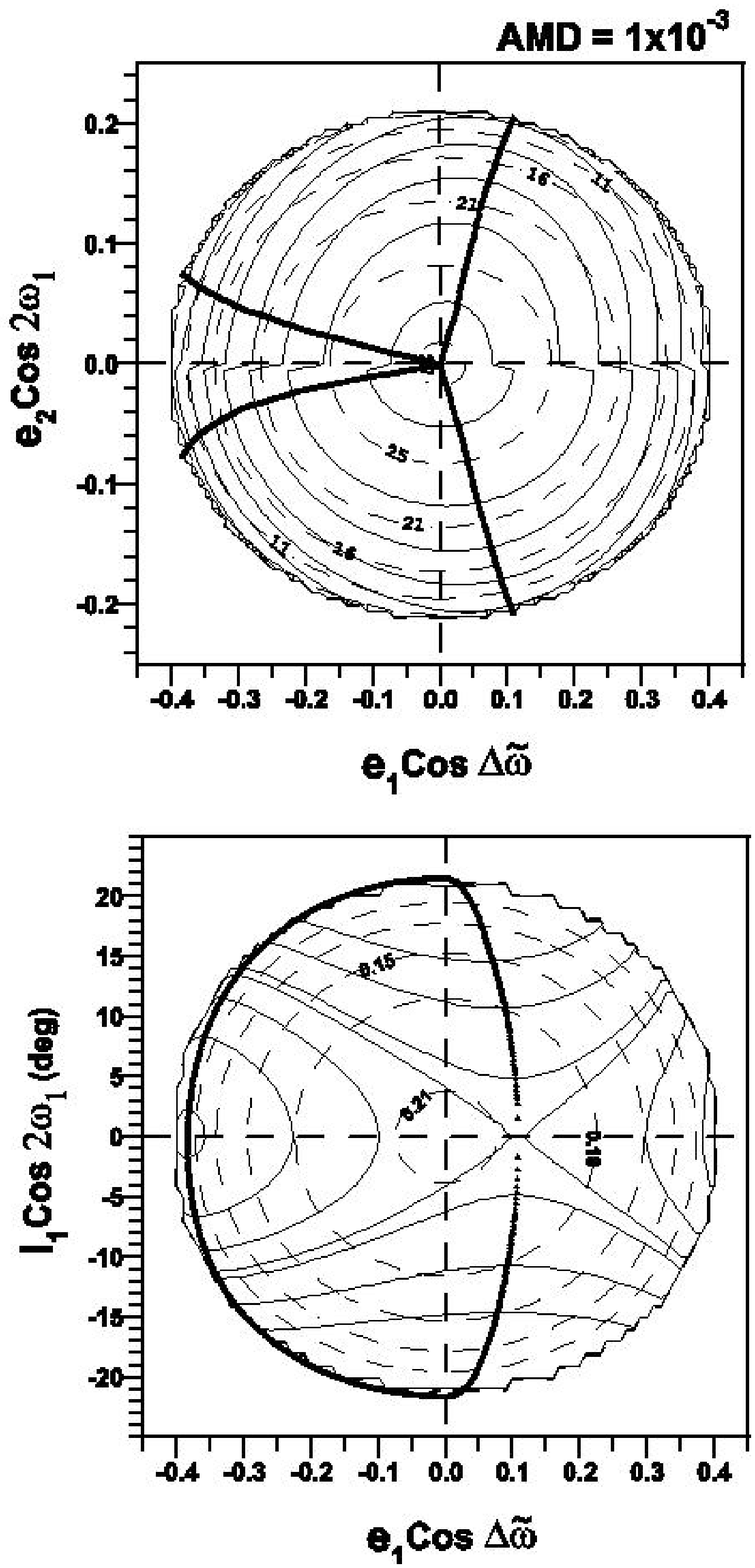}}
\caption{Michtchenko {\it et al.}}
\label{fig:fig5}
\end{figure}

\begin{figure}[ht]
\centerline{\hspace{-1cm}\epsfxsize=15cm\epsffile{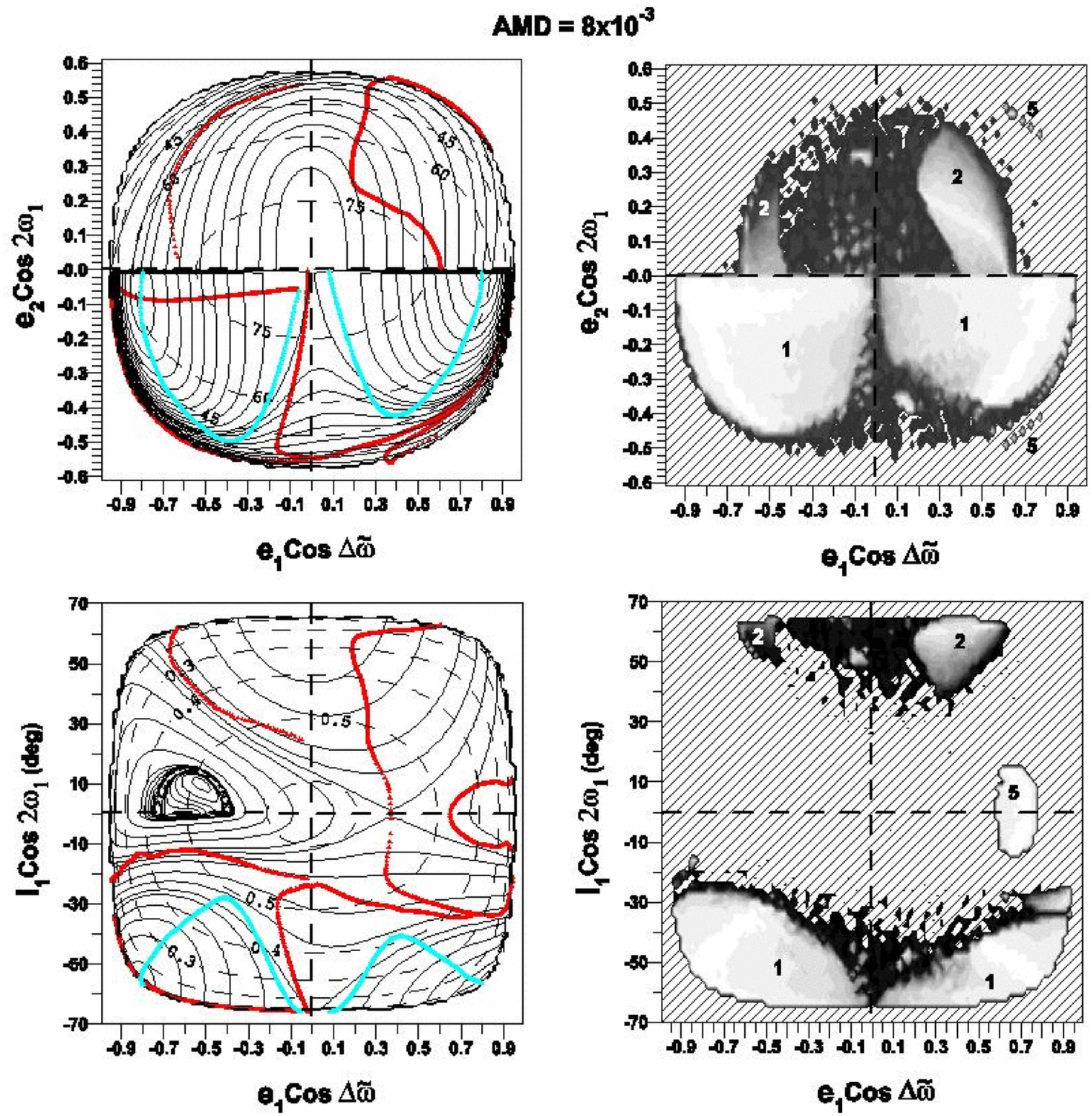}}
\caption{Michtchenko {\it et al.}}
\label{fig:fig6}
\end{figure}

\hspace{-0.5cm}Figure 5. Same as Fig.~\ref{fig:fig2}, but with the condition $\sin\omega_1=0$.
\vspace{0.5cm}

\hspace{-0.5cm}Figure 6. Same as Fig.~\ref{fig:fig3} ({\it left column}), but for ${\rm AMD}=1\times 10^{-3}$. The only periodic solutions of the secular Hamiltonian correspond to the oscillating $\Delta\varpi$.
\vspace{0.5cm}

\hspace{-0.5cm}Figure 7. Same as Fig.~\ref{fig:fig3}, but for ${\rm AMD}=8\times 10^{-3}$. The domain of the secular resonance, where the angle $\Delta\varpi$ in libration around $0^\circ$ and $\omega_1$ in circulation, are marked by {\bf 5}.
\vspace{0.5cm}

\hspace{-0.5cm}Figure 8. Dynamical maps obtained for $I_{\rm mut}=5^\circ$ ({\it top}) and $I_{\rm mut}=15^\circ$ ({\it bottom}). The curves are the location of analytically calculated periodic orbits with $\Delta\dot\varpi=0$. The lighter regions indicate regular oscillation of $\Delta\varpi$ (around $0$ or $180^\circ$), whereas the darker regions indicate its regular circulation. The domains, where planetary collisions occur within 0.5\,Myr, are hatched.
\vspace{0.5cm}

\begin{figure}[ht]
\centerline{\hspace{-1cm}\epsfxsize=8cm\epsffile{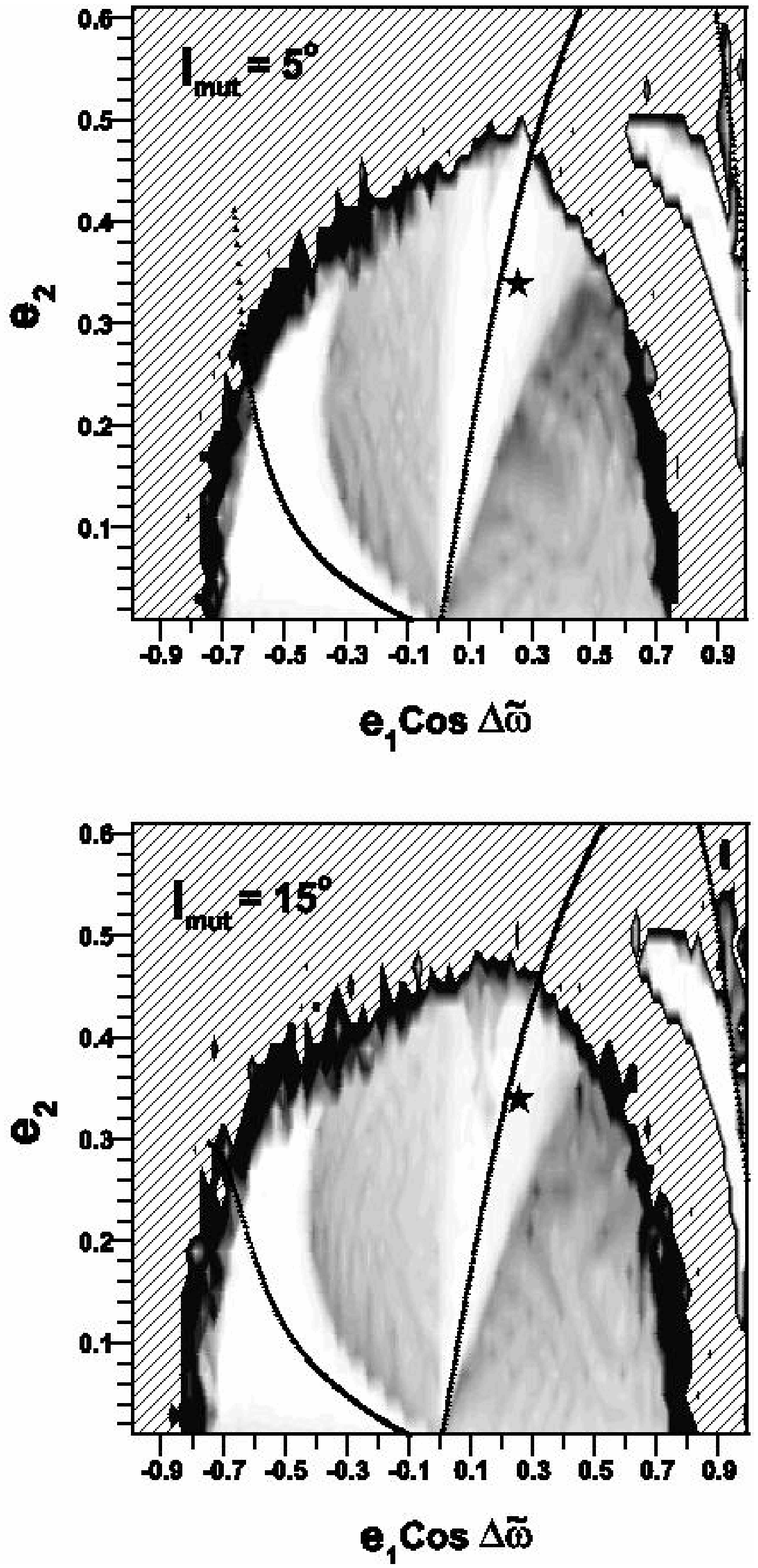}}
\caption{Michtchenko {\it et al.}}
\label{fig:fig7}
\end{figure}

\begin{figure}[ht]
\centerline{\hspace{-1cm}\epsfxsize=8cm\epsffile{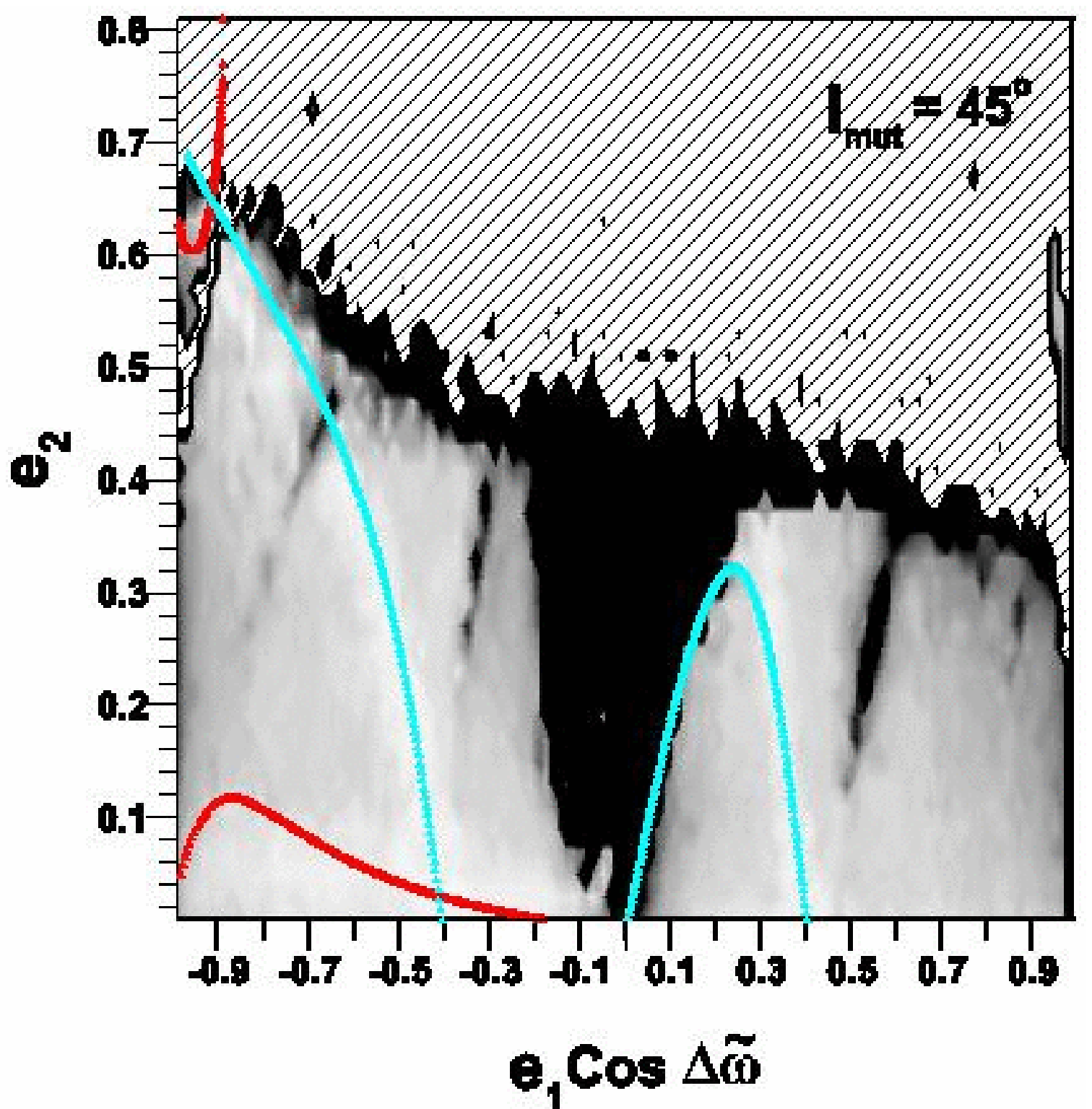}}
\caption{Michtchenko {\it et al.}}
\label{fig:fig8}
\end{figure}

\begin{figure}[ht]
\centerline{\hspace{2.5cm}\epsfxsize=12cm\epsffile{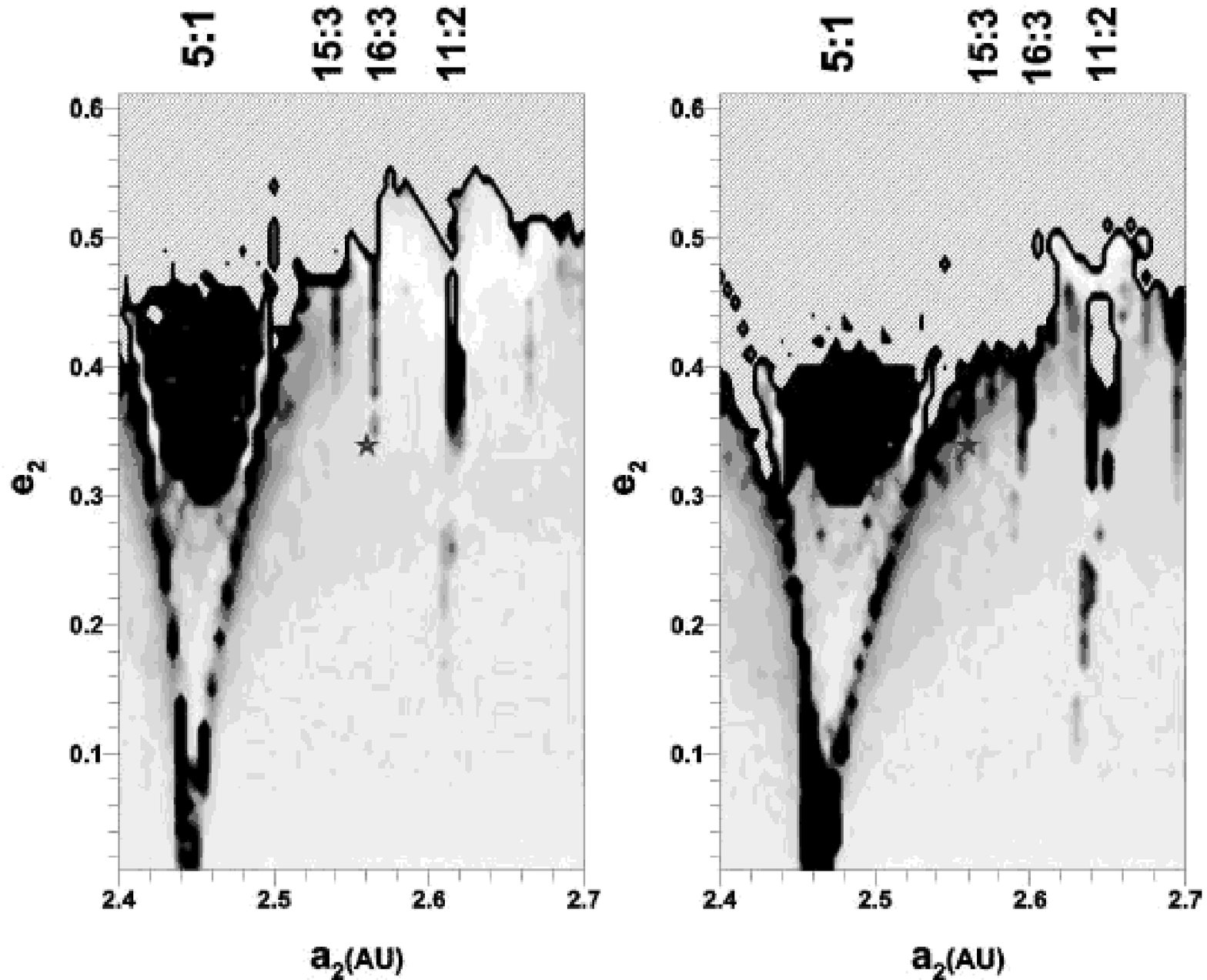}}
\caption{Michtchenko {\it et al.}}
\label{fig:fig9}
\end{figure}

\hspace{-0.5cm}Figure 9. Same as in Fig.~\ref{fig:fig7}, but for $I_{\rm mut}=45^\circ$. The blue curves are the location of analytically calculated periodic orbits with $\dot{\omega}_1=0$, while the red curves of orbits with $\Delta\dot\varpi=0$.
\vspace{0.5cm}

\hspace{-0.5cm}Figure 10. Dynamical maps of the region around the $\upsilon$ Andromedae system on the $(a_2,e_2)$--plane of initial semi-major axis and eccentricity of the outer planet {\bf d}. {\it Left:} Map calculated with planetary masses of the current edge-on system,$m_1=1.83M_{\rm Jup}$ and $m_2=3.97M_{\rm Jup}$; {\it Right:} the same, but with masses $m_1=2\times 1.83M_{\rm Jup}$ and $m_2=2\times 3.97M_{\rm Jup}$. The high-order mean-motion resonances are labeled on the top of the graph. The domains, where planetary collisions occur within 0.5\,Myr, are hatched. The position of the edge-on outer $\upsilon$ Andromedae system is indicated by a star.

\clearpage
\pagebreak \setlength{\unitlength}{1mm}

\clearpage
\pagebreak \setlength{\unitlength}{1mm}

\clearpage
\pagebreak \setlength{\unitlength}{1mm}

\clearpage
\pagebreak \setlength{\unitlength}{1mm}

\clearpage
\pagebreak \setlength{\unitlength}{1mm}

\clearpage
\pagebreak \setlength{\unitlength}{1mm}

\clearpage
\pagebreak \setlength{\unitlength}{1mm}

\clearpage
\pagebreak \setlength{\unitlength}{1mm}

\clearpage
\pagebreak \setlength{\unitlength}{1mm}

\clearpage
\pagebreak \setlength{\unitlength}{1mm}

\end{document}